\documentclass[journal]{IEEEtran}

\ifCLASSINFOpdf
\else
\fi

\usepackage{amsmath, amsfonts}
\usepackage{algorithm}%
\usepackage{algorithmic}
\usepackage{setspace}
\usepackage{array}
\usepackage[caption=false,font=normalsize,labelfont=sf,textfont=sf]{subfig}
\usepackage{textcomp}
\usepackage{fixltx2e}
\usepackage{dblfloatfix}
\usepackage{float}
\usepackage{hyperref}
\usepackage{url}
\usepackage{verbatim}
\usepackage{graphicx}
\usepackage{cite}
\usepackage{bbold}

\usepackage{caption}
\usepackage{textcomp}
\usepackage{xcolor}
\usepackage{amssymb}
\usepackage{mathtools}
\usepackage{color}
\usepackage[utf8]{inputenc}
\usepackage[english]{babel}
\usepackage{amsthm}

\usepackage{comment}
\usepackage{pdfpages}
\usepackage{booktabs}
\usepackage{multirow}

\newcommand{\bF}{\mathbf{F}}
\newcommand{\bG}{\mathbf{G}}
\newcommand{\bT}{\mathbf{T}}
\newcommand{\bI}{\mathbf{I}}

\newcommand{\bu}{\mathbf{u}}
\newcommand{\bv}{\mathbf{v}}

\newcommand{\bd}{\mathbf{d}}

\renewcommand{\comment}[1]{}

\begin{document}
\title{Ptychography using Blind Multi-Mode PMACE}

\author{Qiuchen Zhai,~\IEEEmembership{Student Member,~IEEE,} Gregery T. Buzzard,~\IEEEmembership{Senior Member,~IEEE,} Kevin Mertes, Brendt Wohlberg,~\IEEEmembership{Fellow,~IEEE,} Charles A. Bouman,~\IEEEmembership{Fellow,~IEEE}
\thanks{The work of Q. Zhai was supported by U.S. Department of Energy through LANL. 
The work of C. A. Bouman was supported in part by the U.S.
Department of Energy and the Showalter Trust. 
The work of G.T. Buzzard was partially supported by NSF CCF-1763896. 
The work of B. Wohlberg was supported by the LDRD program of LANL under project number 20200061DR.
The publication release number is LA-UR-24-33356.}

\thanks{Qiuchen Zhai and Charles A. Bouman are with the School of Electrical and Computer Engineering, Purdue University, 465 Northwestern Ave., West Lafayette, IN 47907, USA (qzhai@purdue.edu and bouman@purdue.edu).

Gregery T. Buzzard is with the Department of Mathematics, Purdue University, West Lafayette, IN 47907, USA (buzzard@purdue.edu).

Kevin Mertes is with Physical Chemistry and Applied Spectroscopy Group, Los Alamos National Laboratory, Los Alamos, NM 87545 USA (kmmertes@lanl.gov).

Brendt Wohlberg is with CCS-3 Information Sciences Group, Los Alamos National Laboratory, Los Alamos, NM 87545 USA (brendt@ieee.org).}%

}

\maketitle

\begin{abstract}
Ptychography is an imaging technique that enables nanometer-scale reconstruction of complex transmittance images by scanning objects with overlapping illumination patterns. However, the illumination function is typically unknown, which presents challenges for reconstruction, especially when using partially coherent light sources.

In this paper, we introduce Blind Multi-Mode Projected Multi-Agent Consensus Equilibrium (BM-PMACE) for blind ptychographic reconstruction. We extend the PMACE framework for distributed inverse problems to jointly estimate the complex transmittance image and multiple, unknown, partially coherent probe functions. Importantly, our method maintains local probe estimates to exploit complementary information at multiple probe locations. Our method also incorporates a dynamic strategy for integrating additional probe modes.
Through experimental simulations and validations using both synthetic and measured data, we demonstrate that BM-PMACE outperforms existing approaches in reconstruction quality and convergence rate.
\end{abstract}


\IEEEpeerreviewmaketitle

\section{Introduction}

\IEEEPARstart{P}{tychography} is a lensless imaging technique for non-destructive imaging of samples at nanometer resolutions and can reveal precise information about the sample's density and thickness \cite{nellist1995resolution, rodenburg2004phase, rodenburg2007hard,pfeiffer2018x, rodenburg2019ptychography}. 
Ptychography can be combined with advanced imaging modalities, such as tomography \cite{li2018multi, chang2020ptychographic, batey2022high, gorecki2023ptychographic, pelz2023solving}, to produce high-resolution images and provide detailed insights into complex nano-scale structures. 
As a result, ptychography has applications in fields ranging from biomedical imaging \cite{wang2023optical, guo2023depth} to material science \cite{wang2017electron, jiang2022ptychographic, li20224d}. 

In ptychography, a sequence of overlapping regions is scanned with a coherent or partially coherent light source known as a probe \cite{chang2018partially}.
At each probe location, the intensity of the resulting diffraction pattern is measured by an imaging detector positioned in the far-field plane.
These real-valued diffraction measurements are used to reconstruct the complex transmission image by solving a phase recovery problem. 
Since the phase is lost in measurement, the forward model for ptychography is non-linear, the reconstruction problem is typically ill-posed, and the associated optimization problems are non-convex and non-smooth \cite{chang2023fast}. 
However, the overlapping probe locations provide redundancy in the measurements that make it possible to solve the associated inverse problem.

Accurate reconstruction of the complex transmittance image depends critically on a precise characterization of the light probe \cite{deng2015continuous, chang2019blind, tamaki2024near}. 
However, in real-world experiments, the probe function is rarely fully-known, and typically only partially coherent.
This partial coherence arises because any vibration, energy spread in the beam, or environmental interference can disrupt the coherence and lead to incoherent wave superposition \cite{rodenburg2008ptychography, moxham2020hard}. 
Reconstruction from partially coherent data enables the collection of data with higher flux light sources \cite{yao2020multi} and fly-scan techniques \cite{deng2015opportunities, huang2015fly, pelz2014fly}, both of which can reduce acquisition time.
However, with partially-coherent data, single-mode ptychographic reconstruction may fail or produce poor results with artifacts \cite{thibault2013reconstructing, batey2014information}. 
This motivates a form of ptychography known as blind reconstruction, in which the multi-mode complex probe function and the complex transmittance image are estimated jointly \cite{fannjiang2020blind, hesse2015proximal}.

Table~\ref{tab:comparison_table} lists several existing ptychographic reconstruction algorithms along with relevant characteristics and qualitative performance.
Among the methods that are designed for partially coherent probes, a common approach is to reconstruct multiple mutually non-coherent probe modes using blind multi-mode reconstruction.
Ptychographic engine (PIE) \cite{faulkner2004movable, rodenburg2004phase} and its variants \cite{maiden2009improved, maiden2012ptychographic, maiden2017further}, Difference Map (DM) \cite{thibault2008high, thibault2009probe}, and SHARP \cite{marchesini2013augmented, marchesini2016sharp} use alternating updates of the complex transmittance and the probe function. Both ePIE \cite{yao2021broadband, long2022single} and DM \cite{dong2018high, shi2018multi, fang2020accelerated} have been modified to support blind multi-mode reconstruction.  The multi-mode ePIE algorithm employs a serial approach in which each probe location is updated in sequence.
This has the advantage of speeding per-iteration convergence, since the multi-mode probe estimate is also updated with each image patch update.
However, serial update is not practical for large ptychography problems, which may have millions of probe locations \cite{yu2022scalable}.
Moreover, gradient-based algorithms such as ePIE can sometimes exhibit slow convergence or become trapped in local minima when the data are noisy and/or sparse \cite{bunk2008influence, melnyk2023convergence}.

\begin{table*}[t]
  \small
  \setstretch{1.2}
  \centering
  \begin{tabular}{ |c|c|c|c|c|c|c|c| } 
    \hline
    \rule{0pt}{1.6em}
    Algorithm & Type 
    & \parbox{0.5in}{\centerline{Blind } \centerline{Recon.}}   
    & \parbox{0.5in}{\centerline{Multi-Mode} \centerline{Probe}} 
    & \parbox{0.8in}{\centerline{Performance on} \centerline{sparse data}} 
    & \parbox{0.8in}{\centerline{Performance on} \centerline{noisy data}} \\[1.6ex] \hline
    ePIE~\cite{maiden2017further, long2022single} & Serial & Yes & Yes & Fair & Fair\\
    \hline
    DM~\cite{elser2003phase, dong2018high} & Parallel & Yes & Yes & Good & Good \\ 
    \hline
    SHARP~\cite{marchesini2016sharp} & Parallel & Yes & No & Good & Good\\
    \hline
    WF/AWF~\cite{xu2018accelerated} & Parallel & No & No & Good & Fair \\ 
    \hline
    GDP-RAAR/ADMM~\cite{enfedaque2019high, chang2018partially} & Parallel & Yes & Yes & Good & Good\\ 
    \hline
    PMACE~\cite{zhai2023projected} & Parallel & No & No & Excellent & Excellent \\  
    \hline
    BM-PMACE [this paper] & Parallel & Yes & Yes & Excellent & Excellent \\  
    \hline
  \end{tabular}
\caption{Comparison of ptychography reconstruction algorithm features.}
\label{tab:comparison_table}
\vspace{-1.6pt}
\end{table*}

Alternatively, DM algorithms allow for parallelization and also support blind multi-mode reconstruction.
These methods refine the estimate across multiple scan locations simultaneously, resulting in more generalized estimates. 
This approach improves computational efficiency by leveraging distributed computation and offers better scalability for large-scale datasets. 
However, parallel methods may require additional iterations to converge compared to serial approaches \cite{nashed2014parallel}. 
More recent algorithms include GDP-RAAR \cite{enfedaque2019high} and GDP-ADMM \cite{chang2018partially}.
While both algorithms have been demonstrated to be effective, they both use a non-standard physical model in which the probe convolution occurs after the energy detection.
While GDP-RAAR and GDP-ADMM also support parallel processing and blind multi-mode reconstruction, they may not accurately capture the underlying physics.

Finally, Wirtinger flow (WF) \cite{xu2018accelerated} and Projected Multi-Agent Consensus Equilibrium (PMACE) \cite{zhai2021projected, zhai2023projected} have been proposed with a focus on reconstruction of the complex transmittance image, without providing a well-defined strategy for probe estimation.  However, despite the lack of probe estimation, the PMACE algorithm has been shown to provide more accurate reconstructions than alternative algorithms when the probe locations are sparse and noisy \cite{zhai2021projected, zhai2023projected}.

In this paper, we introduce BM-PMACE, an extension of PMACE that supports blind multi-mode reconstruction of ptychographic data.\footnote{The reference implementation of PMACE is available at \url{https://github.com/cabouman/ptycho_pmace}.}
Since it is based on PMACE, BM-PMACE allows for fast and parallel reconstruction of the transmission image and produces good quality reconstructions with sparse and noisy data.
BM-PMACE builds on PMACE by incorporating a local state consisting of both an image patch and a set of estimated probe modes at each probe location.
This results in an algorithm that is parallelizable for both the image and probe updates, which is critical for practical application on large data sets.
Moreover, we demonstrate that by keeping local probe estimates for each patch, the BM-PMACE algorithm achieves much faster and more robust per-iteration convergence for blind multi-mode reconstruction.

More specifically, we make the following contributions:
\begin{itemize}
\item Formulate the BM-PMACE algorithm for blind multi-mode ptychographic reconstruction based on the PMACE algorithm;
\item Introduce a novel distributed state with a multi-mode probe estimate for each image patch;
\item Introduce a method for automatic addition of probe modes.
\end{itemize}

Our experimental results on both synthetic and measured ptychography data demonstrate that the distributed probe estimates consistently result in faster and more robust per-iteration convergence and better reconstructed image quality than competing methods.

\section{Overview of Blind Multimode Ptychography}

\begin{figure}[ht!]
    \centering
    \includegraphics[width=0.5\textwidth]{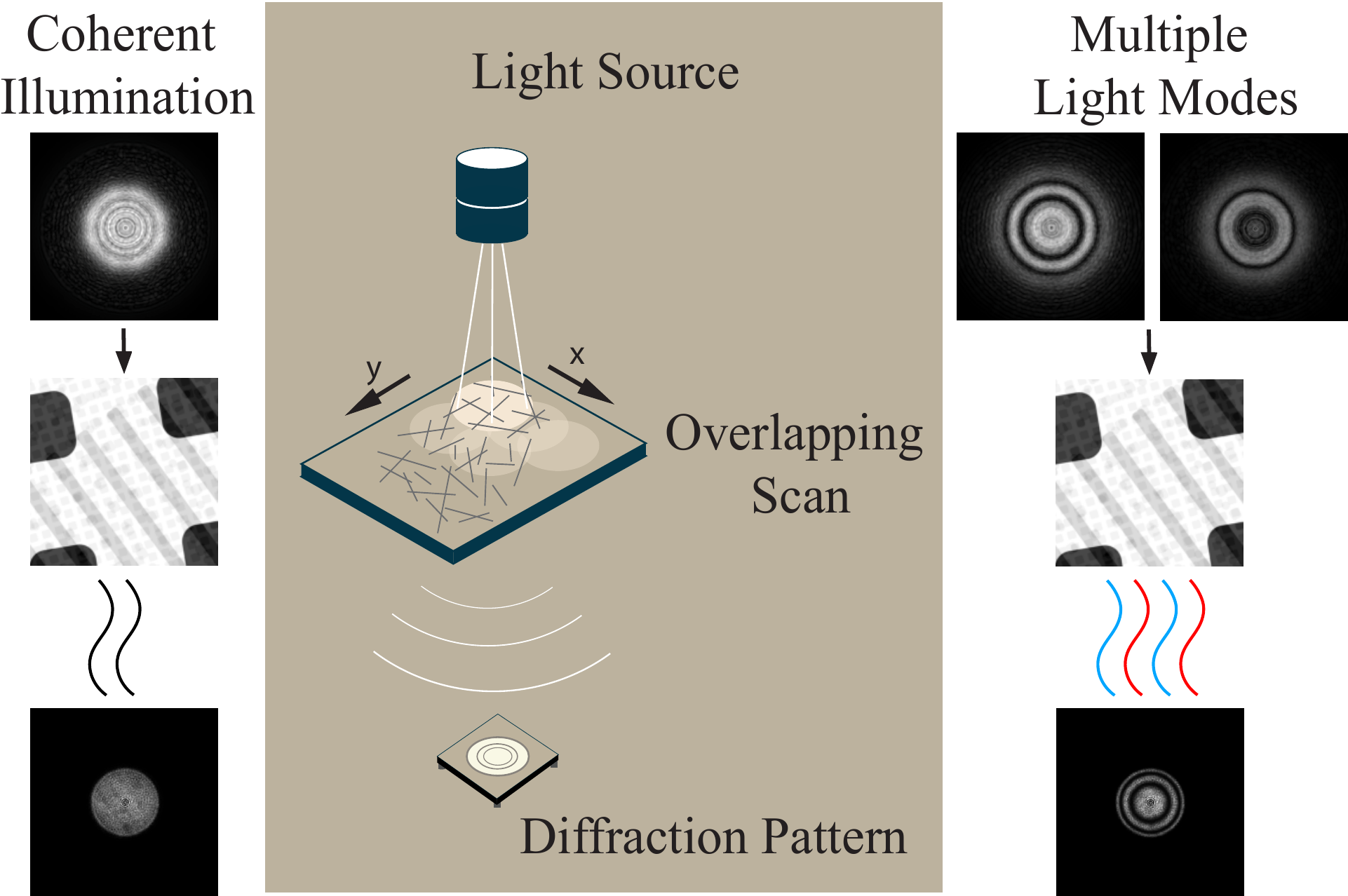}
    \caption{Illustration of ptychography with coherent and partially coherent light sources. On the left, the coherent light source illuminates part of the sample, resulting in a uniform and sharp diffraction pattern. On the right, the partially coherent light source produces a superimposed diffraction pattern.}
    \label{fig:ptychography}
\end{figure}

Figure~\ref{fig:ptychography} illustrates how ptychography is performed.
A coherent radiation source, such as soft X-rays or electrons, is used to generate a ``probe'' that illuminates a series of overlapping patches in a larger flat object.
For each probe location, the intensity of the resulting far-field diffraction pattern is measured. 
The 2D image is then reconstructed by estimating the complex transmittance image as part of a larger phase recovery problem.

To first order, the phase of the reconstructed 2D image changes in proportion to the thickness of the sample and hence gives precise information about sample thickness.  This fact is due to the change in transmission time through the sample as a function of refractive index and thickness.  

In blind multi-mode ptychography, we assume that the illumination source is both unknown and partially coherent. 
In this case, the probe is not fully coherent, so we model the probe as consisting of multiple, mutually incoherent modes. 
These modes are summed in energy, and a 2D complex probe cross section must be estimated for each mode individually.

\subsection{Notation and Forward Model}

We let $x \in \mathbb{C}^{N_{1} \times N_{2}}$ denote the complex transmittance function of the object being imaged, where $N_1$ and $N_2$ represent the number of pixels along the horizontal and vertical dimensions of the object.  We let $\left \{ d_{k} \in \mathbb{C}^{N_{p} \times N_{p}}, k=0, \ldots, K-1 \right \}$ represent the set of multiple probe modes used to illuminate the sample, where $N_p$ represents the number of pixels along each dimension of the probe mode and $K$ denotes the total number of probe modes.
Furthermore, let $D_{k} = \text{diag}(d_{k})$ denote the diagonal matrix representing the complex illumination function of the $k$th probe mode. Then the forward model of the diffraction pattern from the $k$th mode of the $j$th probe location is given by
$$
\bar{I}_{j,k} = | \mathcal{F} D_{k} v_{j} |^2 = | \mathcal{F} D_{k} P_{j} x |^2 \ ,
$$
where $P_j: \mathbb{C}^{N_{1} \times N_{2}} \rightarrow \mathbb{C}^{N_{p} \times N_{p}}$ denotes the linear projection operator that extracts the localized patch associated with $j$th probe location, $v_j = P_j x$ represents the corresponding patch, $D_{k}$ multiplies by the complex probe intensity, and $\mathcal{F}$ denotes the 2D orthonormal discrete Fourier Transform.
Note that the intensity is proportional to the square of the electric field.

The detector measurements $I_{j}$ are then Poisson-distributed random variables with means that are proportional to the sum of the energy from all the modes. 
Since the probe modes are not coherently related, the phases are relatively random and the modes are additive in energy.
Consequently, we have that
\begin{align}
    \label{eq: meas_intensity}
    I_{j} &= \text{Pois} \left ( \sum_{k=0}^{K-1} \bar{I}_{j,k} \right ) \nonumber \\
    &= \text{Pois} \left ( \sum_{k=0}^{K-1} | \mathcal{F} D_{k} v_{j} |^2 \right ) \ ,
\end{align}
where $\text{Pois} \left ( \lambda \right )$ denotes an array of independent Poisson distributed random variables with means parameterized by the array $\lambda$.
We note that GDP-RAAR \cite{enfedaque2019high} and GDP-ADMM \cite{chang2018partially} make the approximation of interchanging the non-linear operator $|\cdot |^2$ and the linear operator $D_k$.

While some methods directly process the Poisson measurements~\cite{bian2016fourier}, we apply a variance-stabilizing transformation to the detector measurements by taking the square root~\cite{godard2012noise}.
Therefore, the forward model for our multi-mode ptychography system is given by
\begin{equation}
    \label{eq: meas_mag}
    y_{j} = \sqrt{ \text{Pois} \left ( \sum_{k=0}^{K-1} | \mathcal{F} D_{k} v_{j} |^2 \right )} \ .
\end{equation}
This transformation simplifies the forward model of our approach.

In the following sections, we extend the PMACE framework of~\cite{zhai2021projected, zhai2023projected} to the case of blind multi-mode reconstruction of ptychography data.
To do this, we first derive the PMACE algorithm for non-blind multi-mode reconstruction in Section~\ref{sec:ImageReconstruction}.
We then derive the corresponding algorithms for PMACE reconstruction of the probe modes in Section~\ref{sec:ProbeReconstruction}.
Finally, in Section~\ref{sec:IntegratedPMACEAlgorithm} we introduce an integrated algorithm for blind, multi-mode PMACE (BM-PMACE) reconstruction.

\section{Image Reconstruction using PMACE}
\label{sec:ImageReconstruction}

\begin{figure}[t]
    \centering
    \includegraphics[width=0.35\textwidth]{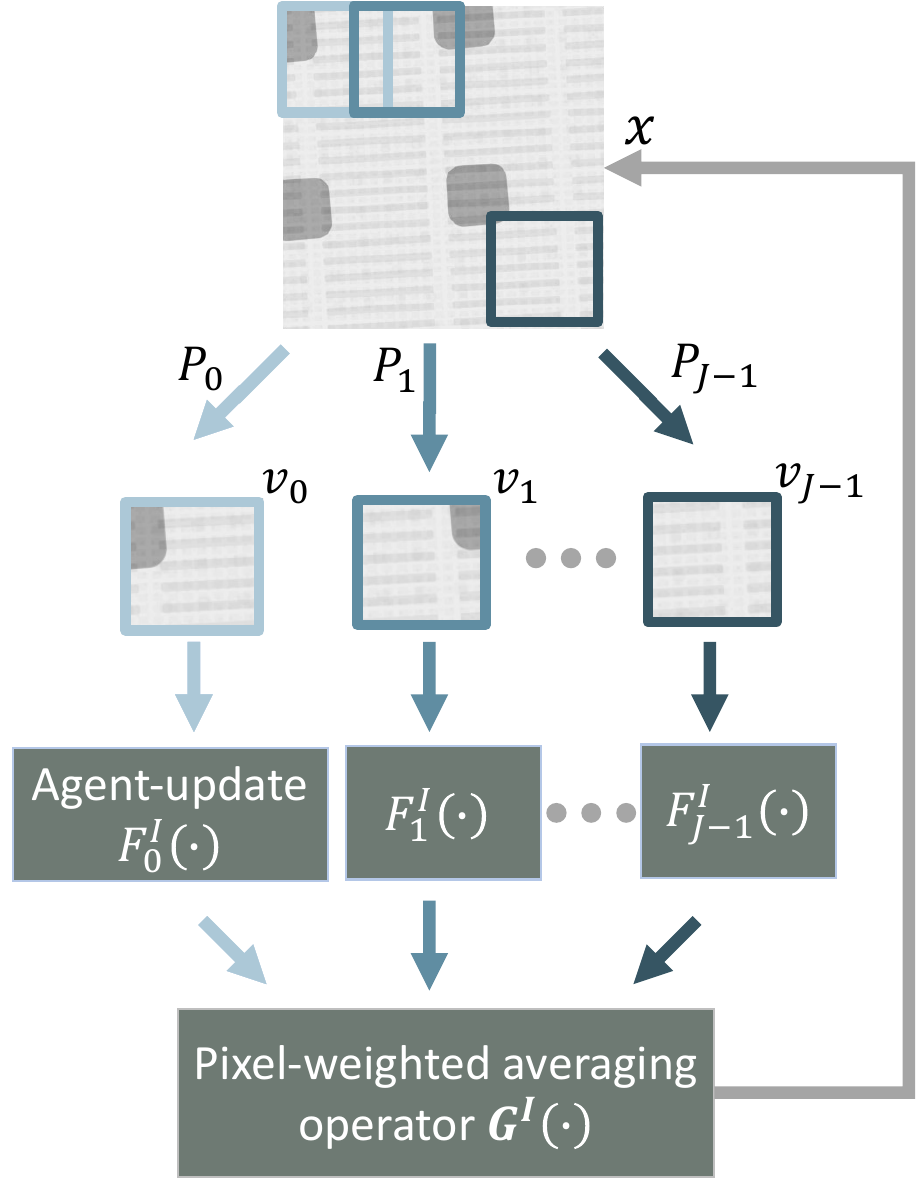}
    \caption{Conceptual overview of the PMACE pipeline for patch refinement.  The state $x$ is divided into overlapping components $v_j$, and distributed to multiple agents $F^I_j (\cdot)$ for local enhancement. These local reconstructions are then combined using a pixel-weighted averaging operator $\bG^I (\cdot)$ to create an integrated global reconstruction, which ensures consistency within overlapping regions.}
    \label{fig:pmace_pipeline_object}
\end{figure}

In this section, we derive a PMACE algorithm for multi-mode reconstruction of the transmission image $x$, assuming that we know the probe modes, $d_k$, $k=0,\ldots K-1$.  We discuss probe estimation in Section~\ref{sec: probe update agent}.  

Figure~\ref{fig:pmace_pipeline_object} illustrates the PMACE pipeline for object refinement and gives a conceptual overview of how the complex transmittance image, patch estimates, data-fitting agents, and the pixel-weighted averaging operator interact to drive the iterative reconstruction process. We give more detail below, but roughly, each patch represents a local region of the object, and updates are performed individually by the agent $F^I_j (\cdot)$ to refine the complex transmittance patches based on the measured diffraction patterns. The pixel-averaging operator $\bG^I (\cdot)$ enforces agreement between overlapping patches by averaging the estimates from the agents. 

\subsection{Image Update Pipeline}

To specify the image update algorithm, recall that $P_j$ is a linear operator that extracts the $j$th patch from the image $x$. 
We can then stack these patch states $v_j = P_j x$ into a larger collection given by
\begin{equation}
\bv = \begin{bmatrix} 
    v_0 \\
    \vdots \\
    v_{J-1} 
    \end{bmatrix} \ .
\end{equation}
Then the PMACE equilibrium for the stacked patches is the solution, $\bv^*$, to the equations
\begin{align} 
\label{eq:pmace-state-eqns}
\bF^I (\bv^*) &= \bG^I (\bv^*) \ ,
\end{align}
where $\bF^I (\cdot)$ and $\bG^I (\cdot)$ are operators introduced below and described in more detail in Section~\ref{sec:PatchUpdateFormula}.

The agent operator $\bF^I (\cdot)$ is a stack of individual agents $F^I_j (\cdot)$ expressed as
\begin{equation}
\bF^I (\bv) = \begin{bmatrix}
    F^I_0 (v_0) \\
    \vdots \\ 
    F^I_{J-1} (v_{J-1})
    \end{bmatrix} \ .
\end{equation}
The agent $F^I_j (\cdot)$ refines the patch estimate $v_j = P_j x$ to make it more consistent with the measured data for the $j$th probe location. 
Notice that $F^I(\cdot)$ depends on the probe modes, $d_k$; however, we suppress this dependency for notational simplicity.

The consensus operator $\bG^I (\bv)$ updates each patch using a weighted average that accounts for the current probe estimates and the patch overlaps in the image.
Importantly, the consensus operator has the property that $\bG^I ( \bG^I ( \bv ))= \bG^I (\bv)$ for all $\bv$.
Intuitively, this property means that averaging twice yields the same result as averaging once.

To solve the PMACE equations~\eqref{eq:pmace-state-eqns}, we reformulate it as a fixed point problem. 
It is shown in \cite{zhai2023projected} that Mann iterations can be applied to iteratively update $\bv$ and converge to a fixed point of $\bT^I = (2 \bG^I - \bI)(2 \bF^I - \bI)$ using 
\begin{equation}  
\label{eq:Mann-T-v}
    \bv \gets (1 - \rho) \bv + \rho \bT^I \bv \ ,
\end{equation}
where $\bI$ denotes the identity operator, and $\rho \in (0,1)$ denotes the Mann averaging parameter. 
The choice of $\rho$ controls the step size toward the fixed point. 
When $\bT^I$ is non-expansive, this iterative calculation guarantees convergence to a solution of the PMACE equation if it has at least one fixed point solution.

\subsection{Image Patch Update Agent}
\label{sec:PatchUpdateFormula}

In this section, we define the agents in $\bF^I(\cdot)$ and $\bG^I(\cdot)$ of Eq.~\eqref{eq:pmace-state-eqns}. 
To derive an expression for $F^I_j(v_j)$, we first observe that if $y_j$ is the measured data, and $v_j$ is a reasonably accurate estimate of the current patch, then Eq.~\eqref{eq: meas_mag} together with an assumption that the Poisson measurement is close to its mean implies that $y_j \approx \sqrt{\sum_{m=0}^{K-1} | \mathcal{F} D_{m} v_{j} |^2 }.$
Hence, for the $k$th mode, we can incorporate the $y_j$ into a new, probe-dependent estimate of the patch by replacing the equality $v_j = D_{k,\epsilon}^{-1} \mathcal{F}^{*} ( \mathcal{F} D_{k} v_{j} )$ with the update 
\begin{align}
\nonumber
\tilde{v}_{j,k} 
&= D_{k,\epsilon}^{-1} \mathcal{F}^{*} \left( 
\frac{y_j}{ \sqrt{\sum_{m=0}^{K-1} | \mathcal{F} D_{m} v_{j} |^2 } } \circ \mathcal{F} D_{k} v_{j} \right) \\
\label{eq:k_mode_patch_update}
&= D_{k,\epsilon}^{-1} \mathcal{F}^{*} \left( y_j \circ 
\frac{\mathcal{F} D_{k} v_{j} }{ \sqrt{\sum_{m=0}^{K-1} | \mathcal{F} D_{m} v_{j} |^2 } }  \right) 
\end{align}
where $\circ$ denotes point-wise multiplication, $\mathcal{F}^{*}$ denotes the inverse (i.e., adjoint) Fourier transform, and $D_{k,\epsilon}^{-1}$ denotes a numerically stable inversion of $D_{k}$.\footnote{We compute the numerically stable inverse of the diagonal entries as $d_\epsilon^{-1} = d^* /(|d|^2 + \epsilon)$ where $\epsilon = 10^{-6}\sqrt{\|d\|^2 /\mbox{dim}(d)}$.}

Using Eq.~\eqref{eq:k_mode_patch_update}, we define the agent $F_j$ as a weighted sum 
\begin{equation}
    \label{eq:Fjv}
    F^{I}_{j} (v_{j}) = (1 - \alpha_{1} ) v_{j} + \alpha_{1} \sum_{k=0}^{K-1} w_{k} \circ \tilde{v}_{j, k}
    \ ,
\end{equation} 
where 
\begin{align} \label{eq: probe weights}
w_{k} &= \frac{ \left \| D_{k} \right \|^{2}_{2} }{ \sum_{m=0}^{K-1} \left \| D_{m} \right \|^{2}_{2} } \ .
\end{align} 
Note that $w_k$ is a scalar that weights the estimates in proportion to their energy, and $\alpha_{1}$ is a parameter that controls the weighting between the previous and new patch estimates. 
We stack the data-fitting agents in Eq.~\eqref{eq:Fjv} to create the data-fitting operator that updates the complex transmittance patches.

The consensus operator $\bG^{I} ( \bv ) =
    \left[
    P_0 \bar{v},
    \ldots,
    P_{J-1} \bar{v}
    \right]^t$
is computed by using $P_j^T$ to position each patch in context in the full image, and forming a weighted average using the probe pixel-wise intensities and the probe energy weights as in Eq.~\eqref{eq: probe weights}.  This gives 
\begin{equation}
\label{eq:consensus_operator_v_in_detail}
    \bar{v} = \sum^{K-1}_{k=0}  w_{k} \Lambda_{k}^{-1}  \sum_{j=0}^{J-1} P_j^T |D_{k}|^\kappa v_j \ , 
\end{equation}
where $\Lambda_{k} = \sum_{j=0}^{J-1} P_{j}^T |D_{k}|^\kappa P_j$, and $1\leq \kappa \leq 2$ is a parameter that effectively controls the weighting as a function of the probe strength.
It is easily shown that $\bG^{I}$ has the required property that $\bG^{I} (\bG^{I} ( \bv )) = \bG^{I}( \bv )$.
Intuitively, $\bG^{I}$ results in a pixel-weighted and mode-weighted average of the patches. 
The solution $\bv^*$ is then computed using iterative updates as shown in Eq.~\eqref{eq:Mann-T-v}.

\section{Probe Reconstruction using PMACE}
\label{sec:ProbeReconstruction}

A key feature of BM-PMACE is the use of separate probe estimates at each probe location and a consensus operator that produces a single probe estimate from these individual estimates.  This approach makes efficient use of the complementary information inherent in patches with different transmission images while also producing a single estimate that informs each individual estimate.  

\begin{figure}[ht]
    \centering
    \vspace{0.3cm}
    \includegraphics[width=0.5\textwidth]{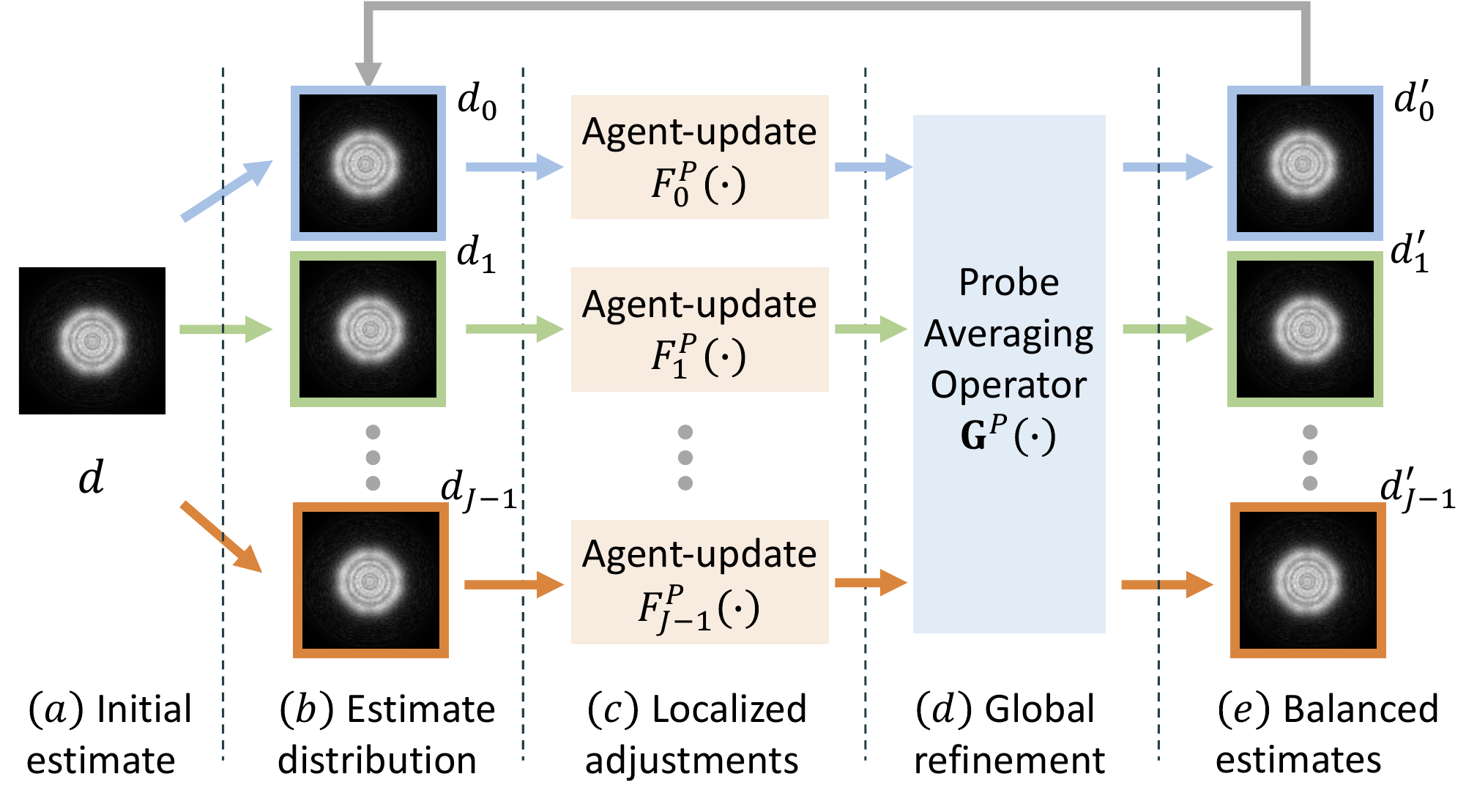}
    \caption{Illustration of the PMACE pipeline for multi-mode probe estimation. 
    The pipeline starts by distributing the global probe estimate $d$ to multiple agents. 
    Each probe agent then performs a localized adjustment using its own update function $F^{P}_{j} \left ( \cdot \right )$.
    The locally refined probe estimates are then averaged by $\bG^{P} \left ( \cdot \right )$ to form an array of identical estimates.
    }
    \vspace{0.3cm}
    \label{fig: PMACE_pipeline_probe}
\end{figure}

Figure~\ref{fig: PMACE_pipeline_probe} illustrates the PMACE pipeline for probe refinement. An initial probe estimate is distributed to multiple agents, each associated with a specific scan position. 
Each agent makes adjustments to its probe estimate using the associated patch transmittance estimate and corresponding measurements.
As in the pipeline in Figure~\ref{fig:PMACE_pipeline_probe}, these local estimates are then averaged to update the global probe, ensuring consistency between different scan locations. 

Intuitively, this algorithm maintains local and global probe mode estimates through $d_{j,k}$ and $\bar{d}_k$, respectively.
The local estimate, based on the characteristics of the reconstructed image and data in a patch, allows the probe modes to adapt to specific variations encountered at each scan location.
The global estimate, formed by the average of these local estimates, averages out local variations and is used to update the transmission image.
In the results section, we show that by keeping this combination of local and global probe estimates, our algorithm achieves fast and robust convergence to a high-quality reconstruction.

\subsection{Probe Update Pipeline} 

In this section, we outline the PMACE algorithm used to estimate the probe modes assuming a known transmission image, $x$.  In Sec.~\ref{sec: probe update agent}, we describe the local probe update agent, and in Sec.~\ref{sec:IntegratedPMACEAlgorithm} we combine probe and patch estimation.  

For probe estimation, we define the collection of probe modes indexed by $k$, which is formed from the estimates at each of $J$ probe locations to yield
\begin{equation}
\bd_{k} = \begin{bmatrix} 
    d_{0,k} \\
    \vdots \\
    d_{J-1, k}
    \end{bmatrix} \ .
\end{equation}
For the estimate of the $k$th mode at the $j$th location, we define an associated agent $F^{P}_{j,k}$ (described in Sec.~\ref{sec: probe update agent}) that updates the corresponding probe estimate. 
This allows the local probe updates to be performed independently based on the associated measurements and patch transmittance.
We stack these individual updates to form the operator
\begin{equation}
\bF_k^P (\bd_k ) = \begin{bmatrix}
    F_{0,k}^P (d_{0,k}) \\
    \vdots \\ 
    F_{J-1,k}^P (d_{J-1,k})
    \end{bmatrix} \ .
\end{equation}
Each $F_{j,k}^P$ depends on the patch $x_j$, but we suppress this dependence for simplicity.  

We define the consensus operator for the probes to be a simple average of the local probe mode estimates given by 
\begin{equation}
\bG^{P} ( \bd_{k} ) = \begin{bmatrix}
    \bar{d}_k \\
    \vdots \\ 
    \bar{d}_k
    \end{bmatrix} \ ,
\end{equation}
where $\bar{d}_k =  \frac{1}{J} \sum^{J-1}_{j=0} d_{j,k}$.

As before, we find the PMACE solution for each probe mode as the fixed point of the operator $\bT_k^P = (2 \bG_k^P -\bI)(2 \bF_k^P - \bI)$ using Mann iterations:
\begin{equation}  
\label{eq:Mann-T-d}
    \bd_k \gets (1 - \rho) \bd_k + \rho \bT_k^P \bd_k \ .
\end{equation}

\subsection{Local Probe Update Agent} \label{sec: probe update agent}

For the local probe update, we need to incorporate the existing probe estimate, $d_j$, with the patch measurements, $y_j$, and the patch transmittance estimate, $P_j x$.  We multiply the probe mode by the transmittance, take the Fourier transform, multiply by the measured values normalized by the full probe amplitude, then take the inverse Fourier transform and divide by the patch to get the new probe mode update.  This gives 
\begin{equation}
    \label{eq:probe_data_fitting_point}
    \tilde{d}_{j, k} = X_{j,\epsilon}^{-1} \mathcal{F}^{*} \left (y_{j} \circ \frac{ \mathcal{F} X_{j} d_{j, k}}{\sqrt{\sum_{m=0}^{K-1} |\mathcal{F} X_{j} d_{j, m}|^2 }} \right ) \ ,
\end{equation}
where $X_{j} = \text{diag}(P_{j} x)$ is a diagonal matrix representing the patch transmittance and $X_{j,\epsilon}^{-1}$ denotes the stable inverse of $X_j$ with $\epsilon$ added for numerical stability. 
The full update for $d_{j,k}$ is then
\begin{equation}
    \label{eq:Fjd}
    F^{P}_{j,k} (d_{j,k}) = (1 - \alpha_{2} ) d_{j,k} + \alpha_{2} \tilde{d}_{j, k} \ ,
\end{equation}
where $\alpha_{2}$ controls the balance between the current estimate of the mode and the new estimate.

Note that our probe update agent uses all probe modes $\tilde{d} \left ( d_{j, *} \right )$, along with the current image patches, $v_j = P_j x$, in its update.
However, each agent refines one specific mode at a time. 
For clarity, we suppress this dependency in Eq.~\eqref{eq:Fjd}. 
The probe data-fitting agent makes localized adjustments that are specific to the characteristics of the data collected at that position.

\section{Integrated PMACE Algorithm for Blind Multi-mode Reconstruction}
\label{sec:IntegratedPMACEAlgorithm}

In this section, we integrate the PMACE algorithms for reconstruction of the transmission image, $x$ and the probe modes, $d_k$, for $k=0,\ldots K-1$.
We first present the pseudo-code for implementation, followed by an explanation of the initialization process and the adaptive mode addition strategy.

Algorithm~\ref{alg: PMACE} provides the pseudo-code for computing the BM-PMACE solution. 
The algorithm starts with an initial estimate of the complex transmittance image $x^{(0)}$ and an initial estimate of a single complex probe function $d^{(0)}$. 
The number of probe modes, $K$, is initialized to 1. 
We set $\rho=0.5$ as the default value for the Mann averaging parameter, which provides a practical balance between convergence speed and stability. 
The probe weight parameter is set as $\kappa=1.25$ since we find that it provides fast convergence speed and high reconstruction quality.

The initial vectors $\mathbf{w}$ and $\mathbf{v}$ are created from the projections of $x^{(0)}$, and the vectors $\mathbf{r}_{0}, \mathbf{s}_{0}$ and $\mathbf{u}_{0}$ are created from $d^{(0)}.$ 
Within each iteration, the algorithm performs updates to the complex transmittance image and probe function alternatively and the main loop continues until convergence is achieved. Note that $\bF^I (\cdot)$ and $\bF^P_k (\cdot)$ both depend on all current probe modes denoted by $\bu_*$. 
This implies that image patches are updated using all current probe modes, and each probe mode is then updated using the new patch estimates and the other probe modes,
and $\bF^P_k (\cdot)$ also depends on the current transmission estimate denoted by $\mathbf{z}$.
Additional modes are added if necessary for better reconstruction quality.
The final reconstruction is obtained by computing $\hat{x}$ from the weighted average of the complex transmittance patches, $v_j$.

\begin{algorithm}[H]
\setstretch{1.1}
\caption{BM-PMACE algorithm.}
\begin{algorithmic}
 \renewcommand{\algorithmicrequire}{\textbf{Input:}}
 \renewcommand{\algorithmicensure}{\textbf{Output:}}
 \REQUIRE Initialization: $x^{(0)} \in \mathbb C^{{N_1} \times N_{2}}$, $d^{(0)} \in \mathbb{C}^{N_{p} \times N_{p}}$, $K=1$ \\ $\qquad$  Design parameters: $\kappa=1.25$, $\rho=0.5$
 \ENSURE  Final Reconstruction: $\hat{x} \in \mathbb C^{{N_1} \times N_{2}}$; $\hat{d}_{k} \in \mathbb C^{{N_p} \times N_{p}}$
 
  \STATE $\mathbf{w} = \mathbf{z} = \mathbf{v} = [v^{(0)}_{0},\dots, v^{(0)}_{J-1}],\; \text{where} \; v^{(0)}_{j} = P_{j} {x}^{(0)} $ \\
  \STATE $\mathbf{r}_{0} = \mathbf{s}_{0} = \mathbf{u}_{0} = [d^{(0)}_{0},\dots, d^{(0)}_{J-1}],\;  $ \\
  
 \WHILE {not converged}
     \STATE \textit{// Update the image patches}
     \STATE $\mathbf{w} \gets \bF^{I}(\mathbf{v}; \mathbf{u}_{*})$
     \STATE $\mathbf{z} \gets \bG^{I}(2\mathbf{w} - \mathbf{v};  \mathbf{u}_{*})$
     \STATE $\mathbf{v} \gets \mathbf{v} + 2\rho (\mathbf{z} - \mathbf{w})$
     
     \STATE \textit{// Update K probe modes}
     \FOR{$k = 0$ to $K-1$}
         \STATE $\mathbf{r}_{k} \gets \bF_k^{P}(\mathbf{s}_k; \mathbf{s}_{*}, \mathbf{z})$
         \STATE $\mathbf{u}_{k} \gets \bG^{P}(2\mathbf{r}_{k} - \mathbf{s}_{k})$
         \STATE $\mathbf{s}_{k} \gets \mathbf{s}_{k} + 2\rho (\mathbf{u}_{k} - \mathbf{r}_{k})$
     \ENDFOR
     \STATE \textit{// Adaptive mode addition}
     \IF{more modes needed}
         \STATE Add mode using Eq. \eqref{eq: add_mode}
         \STATE $K \gets K + 1$
     \ENDIF
 \ENDWHILE
 
 \RETURN $\hat{x} \gets \sum^{K-1}_{k=0}  w_{k} \Lambda_{k}^{-1}  \sum_{j=0}^{J-1} P_j^T |D_{k}|^\kappa v_j$; \\  $\qquad$  $\ $ $ $ $\hat{d}_* \gets \sum_{j=0}^{J-1} s_{j, *}$ \\
\end{algorithmic}
\label{alg: PMACE}
\end{algorithm}

Notice that the BM-PMACE algorithm interlaces updates associated with the image patch agent, $\bF^I$, and each probe mode agent, $\bF_k^P$.
These updates are performed in sequence, which aids in fast convergence, but each agent allows parallel updates at each location of the probe.
We note that standard proofs of mathematical convergence no longer apply with interlaced updates.
However, in practice, we have empirically observed that the algorithm converges robustly as demonstrated in Section~\ref{sec:ExpResults}.

\subsection{Initialization Method}
\label{sec:initialization}

Initialization is a critical step in ptychographic reconstruction that can significantly impact the reconstruction speed, quality, and robustness, particularly when dealing with large datasets or noisy datasets. 

We start by initializing $K=1$ and setting the initial probe mode to
\begin{equation}
\label{eq:initial_probe}
    d^{(0)} \gets U_{\eta, z, \triangle_x} \left\{ \frac{1}{J}\sum_{j=0}^{J-1} \left( P_{j} \mathbf{1} \right)^{-1} \mathcal{F}^{*} y_{j} \right\} \ ,
\end{equation}
where $\mathbf{1}$ is a vector of ones, $U_{\eta, z, \triangle_x}$ denotes Fresnel propagation as a function of source wavelength $\eta$, propagation distance $z$ and sampling rate $\triangle_x$. 
Our practical experience has shown that introducing some phase information through the Fresnel propagator facilitates efficient reconstruction and fast convergence to favorable results.
Next, we initialize the transmittance image using the formula:
\begin{equation}
\label{eq:initial_object}
    x^{(0)} \gets \Lambda_{0}^{-1} \sum_{j=0}^{J-1} P_j^T \left(  \frac{ \| y_j \| }{ \| d^{(0)} \| } \mathbf{1}  \right) \ ,
\end{equation}
where $\Lambda_{0} = \sum_{j=0}^{J-1} P_{j}^T P_j $ and $\mathbb{1}$ is a column vector of 1s. This approach ensures the initialized images match the strength of the data collected at different scan locations.

\subsection{Adding Probe Modes}

To perform reconstruction with multiple modes, we first run reconstruction algorithms with initialized images for an adequate number of iterations. 
Subsequent to these iterations, we incorporate the additional modes as needed. 
We then add additional probe modes through the following calculation
\begin{equation}
\label{eq: add_mode}
d_{K} \gets U_{\eta, z, \triangle_x} \left \{ \frac{1}{J} \sum_{j=0}^{J-1} X_{j}^{-1} \left [ \mathcal{F}^{*} \sqrt{\textup{max} \left ( 0, I_{\text{res}} \right )} \ \right ] \right \} \ ,
\end{equation}
where $I_{\text{res}} = I_{j} - \sum_{k=0}^{K-1} \left | \mathcal{F} D_{k} x_{j} \right |^2$ calculates the residual intensity, the $\textup{max}(\cdot)$ function ensures the argument of the square root function is non-negative.

After that, we scale the energy of all the probes to ensure that the total energy matches the original total energy prior to the incorporation of the new mode. 
This step is critical to maintaining a consistent energy distribution across the probes and preventing a single mode from dominating the reconstruction process.

\section{Experimental Results}
\label{sec:ExpResults}

In this section, we present the results of experiments using both synthetic and measured data\footnote{ The code for reproducing experimental results is available at \url{https://github.com/cabouman/ptycho_pmace_papers}. }.. 
We introduce our data simulation process and show experiment results using both single and multiple probe modes for different light sources. 
Our findings demonstrate that BM-PMACE improves the quality of reconstructions and the robustness of convergence across a range of imaging scenarios.

In all experiments, with both synthetic and measured data, we used $\rho=0.5$, $\kappa = 1.25$, and $\alpha_2 = 0.6$.  For the single-mode cases, we used $\alpha_1=0.6$, and in the two-mode cases, we used $\alpha_1 = 0.5$.

\subsection{Single-Mode Blind Reconstruction: Synthetic Data}
\subsubsection{Single-Mode Data Simulation}

\begin{figure}[h!]
    \centering
    \includegraphics[width=0.4\textwidth]{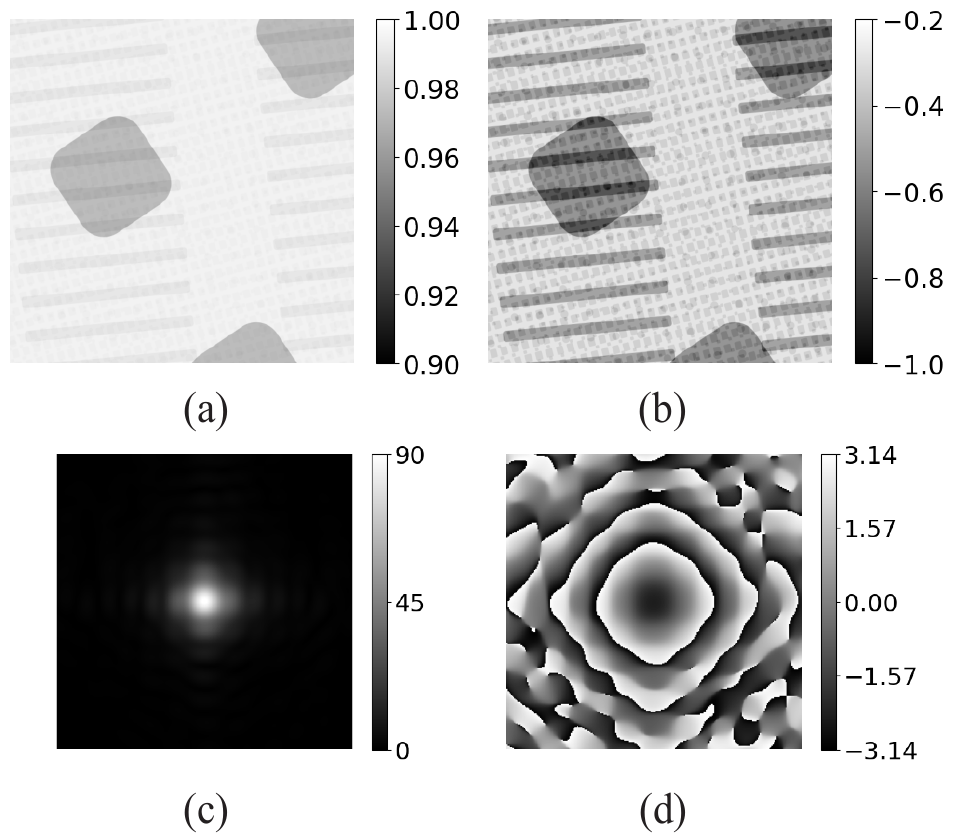}
    \caption{Ground truth image and probe function: (a) Magnitude of the complex ground truth object; (b) Phase of the ground truth object in radians; (c) Magnitude of the complex probe function; (d) Phase of the probe function in radians.}
    \label{fig:ground_truth_single_mode}
\end{figure}

The ground truth image is complex-valued and consists of an $800 \times 800$ array of pixels that models the transmission characteristics of a 5-layer composite material.
The ground truth probe used for sampling the object image was simulated with a photon energy of 8.8 keV and has dimensions of $256 \times 256$ pixels. 
We show the plot of the ground truth images and probes in Figure~\ref{fig:ground_truth_single_mode}.

We used the following formula to simulate the measurements for the $j^\text{th}$ probe location
\begin{equation}
\label{eq: synthetic_data_sim}
    \hat{y}_{j} \gets \sqrt{      
\mathrm{Pois}\left ( r_{p} \frac{I_{j}}{\mathrm{max}_i(\|I_{i}\|_\infty)}  + \lambda \right ) }  \ ,
\end{equation}
where $I_j = \sum_{k=0}^{K-1} | \mathcal{F} D_{k} P_{j} x |^2$, $\|\cdot\|_\infty$ denotes the infinity norm, and $\mathrm{max}_i(\cdot)$ denotes the maximum value over all $i$.
For our simulation of coherent data, we assumed a photon detector with 14-bit dynamic range and the presence of a half-bit of dark current, using $K=1$, $r_{p}=10^4$ and $\lambda = 0.5$. 

We generate random probe locations on a rectangular grid separated by a nominal distance of 36 pixels, but with random offsets uniformly chosen within the range [-5, 5] pixels for each point. 
This approach resembles practical ptychographic experiments and helps avoid periodic reconstruction artifacts \cite{maiden2009improved}. 
This resulted in a probe overlap ratio of $r_{ovlp} \approx 44\%$ using the definition of probe overlap ratio defined in \cite{zhai2023projected}.

\begin{figure}[t]
    \centering
    \includegraphics[width=0.5\textwidth]{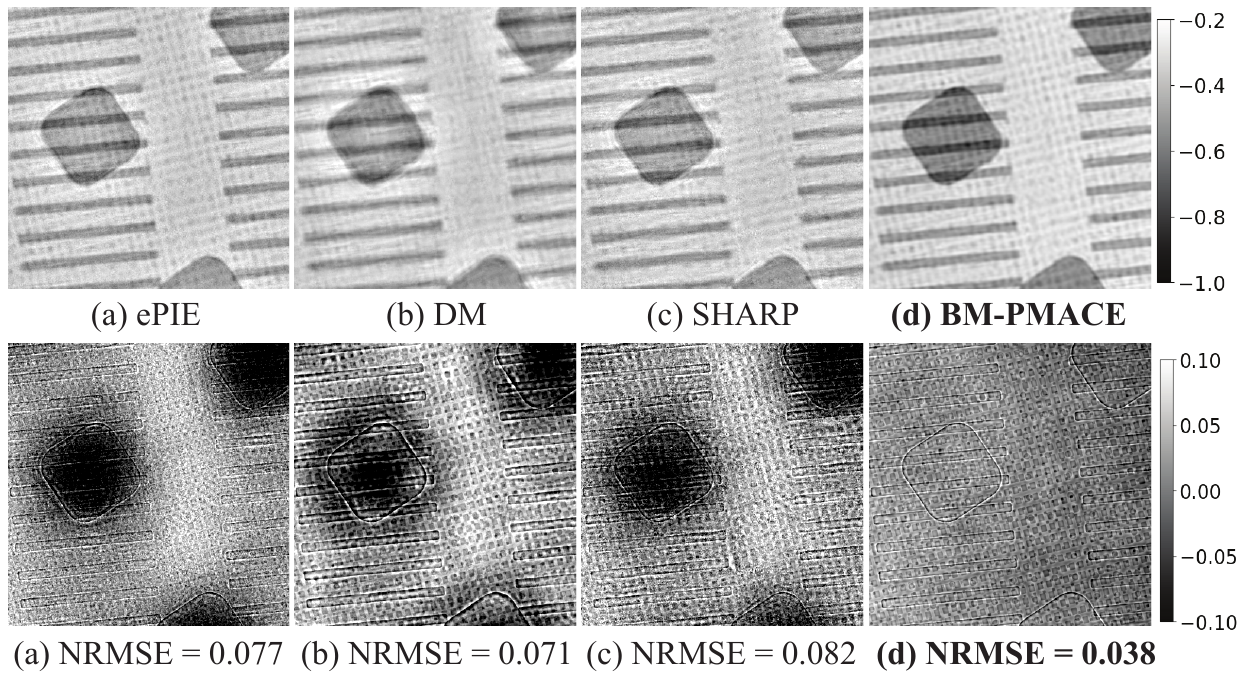}
    \caption{Top:  {\bf Phase} (in radians) of the reconstructed complex transmittance images in Figure~\ref{fig:ground_truth_single_mode} from synthetic single-mode data.  Bottom: Difference between the reconstructed and ground truth phase, with NRMSE values indicated in the subcaptions. }
    \label{fig:single_mode_recon_object_phase}
\end{figure}

\begin{figure}[h!]
    \centering
    \includegraphics[width=0.5\textwidth]{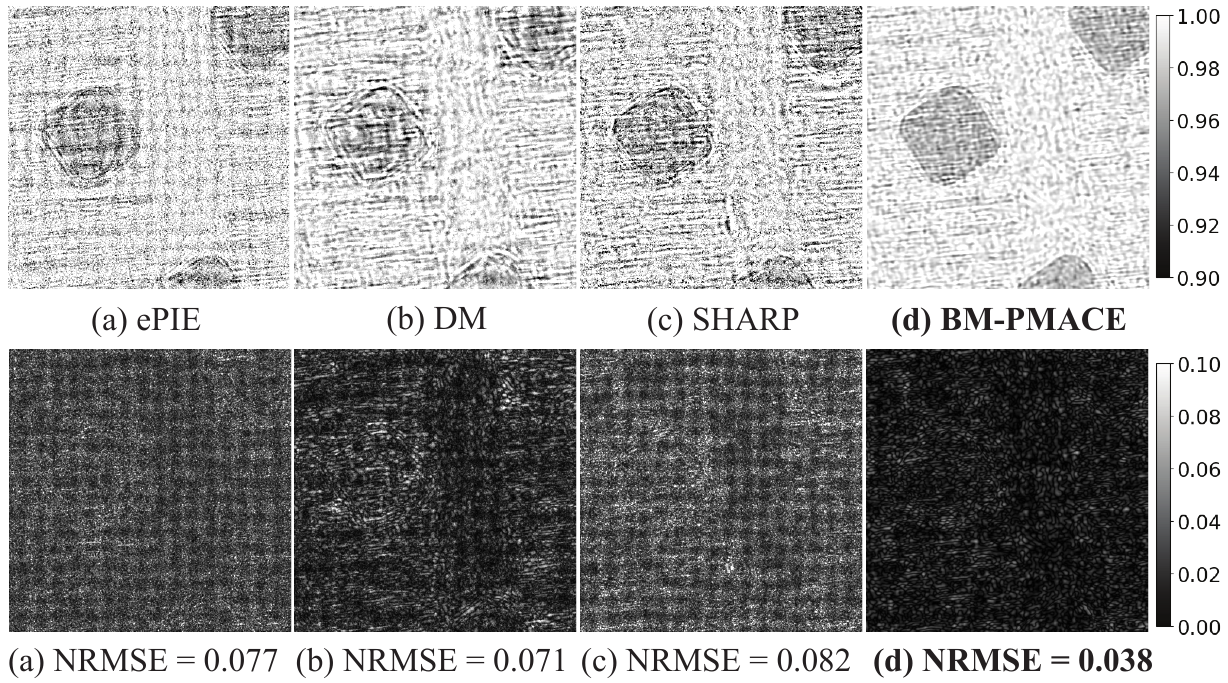}
    \caption{Top:  {\bf Magnitude} of the reconstructed complex transmittance images in Figure~\ref{fig:ground_truth_single_mode} from synthetic single-mode data. Bottom: Amplitdues of error between the complex reconstructions and ground truth. }
    \label{fig:single_mode_recon_object_mag}
\end{figure}

\subsubsection{Single-Mode Reconstruction Results}

Blind reconstructions were performed using the ePIE, AWF, SHARP, and BM-PMACE approaches. 
We optimized the algorithmic parameters for each method using grid search and ran each method for 100 iterations. 
To quantitatively assess the reconstruction quality, we use the normalized root mean square error (NRMSE) value between reconstructed image $\hat{x}$ and the ground truth image $x$ using 
\begin{equation}
    \begin{aligned}
         NRMSE &= \min_{c \in {\mathbb C}\setminus \{0\}} \frac{\| c\hat{x}- x \|}{\| x \|} \ ,
    \end{aligned}
\end{equation}
where $c$ accounts for possible nonunit gain and an unknown phase shift (since ptychographic measurements are invariant to a constant phase shift in the image domain).

Figures~\ref{fig:single_mode_recon_object_phase}
and \ref{fig:single_mode_recon_object_mag} show the reconstructed phases and magnitudes for the transmittance image obtained using ePIE, DM, SHARP, and BM-PMACE.
The NRMSE values are included in the bottom caption of each image. 
In each case, BM-PMACE produced substantially lower reconstruction error than the alternative methods.

\begin{figure}[h!]
    \centering
    \includegraphics[width=0.5\textwidth]{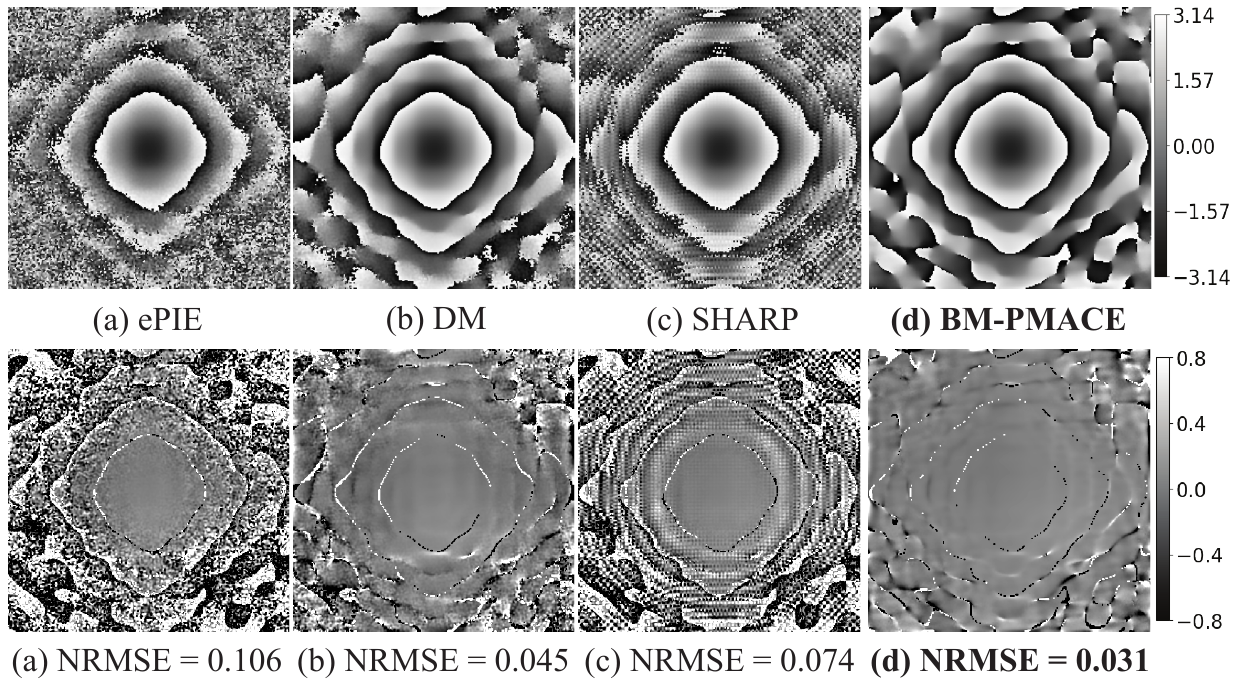}
    \caption{Top:  {\bf Phase} (in radians) of the reconstructed probes from synthetic single-mode data. Bottom: Difference between the reconstructed and ground truth phase, with NRMSE values indicated in the subcaptions.}
    \label{fig:single_mode_recon_probe_phase}
\end{figure}

\begin{figure}[h!]
    \centering
    \includegraphics[width=0.5\textwidth]{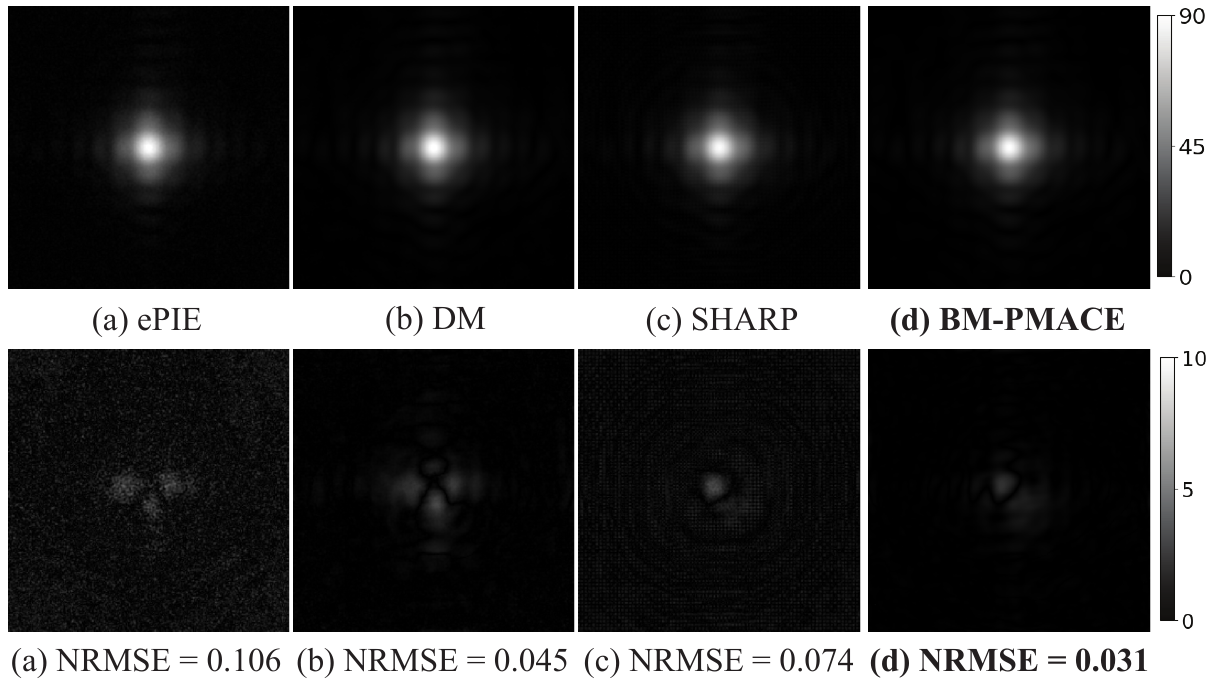}
    \caption{Top:  {\bf Magnitude} of the reconstructed probes from synthetic single-mode data. Bottom: Amplitdues of error between the complex reconstructions and ground truth. }
    \label{fig:single_mode_recon_probe_mag}
\end{figure}

Figures~\ref{fig:single_mode_recon_probe_phase}
and~\ref{fig:single_mode_recon_probe_mag} show the reconstructions and reconstruction error for both the phase and magnitude of the probe resulting from ePIE, DM, SHARP, and BM-PMACE.
The reconstructed phase using ePIE exhibits significant noise, while DM reconstruction mainly shows noise in the border region. 
The SHARP probe contains artifacts near the edges.
Again, the BM-PMACE reconstructions have substantially lower reconstruction error than the alternative methods.

\begin{figure}[!h]
    \centering
    \includegraphics[width=0.4\textwidth]{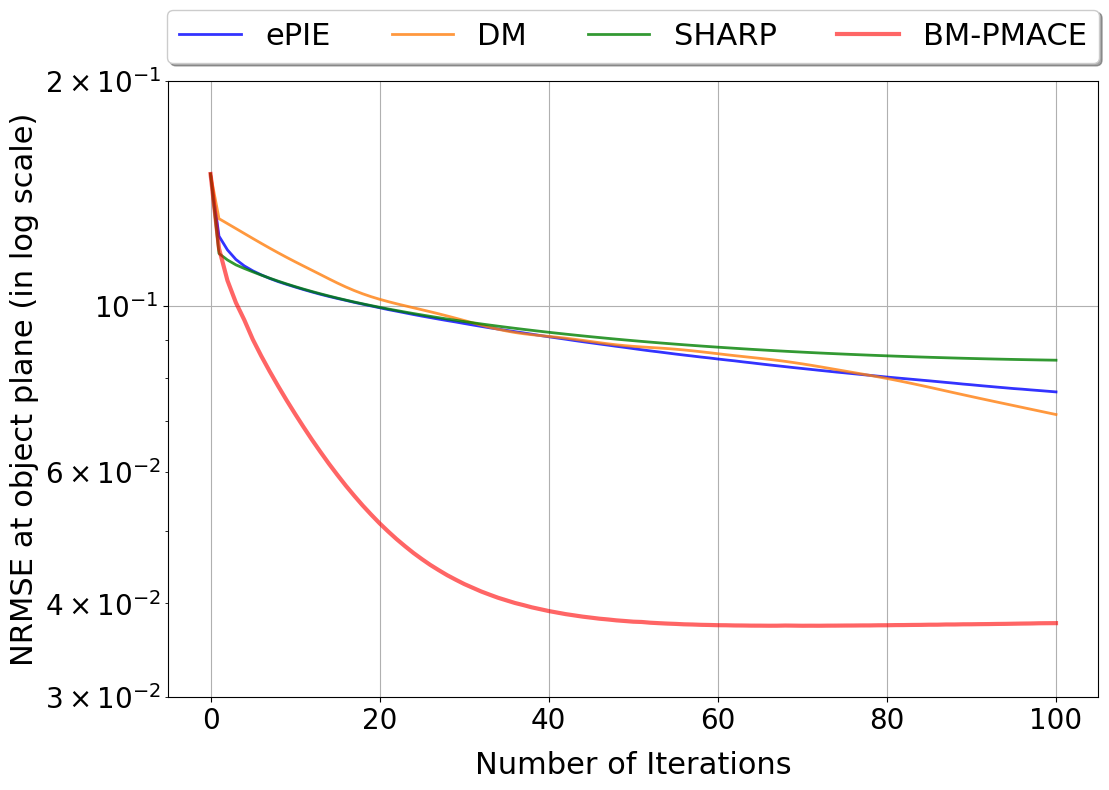}
    \caption{NRMSE between reconstructed object images and ground truth as a function of the number of iterations on synthetic single-mode data with probe overlap ratio $44\% $.}
    \label{fig:single_mode_convergence}
\end{figure}

Figure~\ref{fig:single_mode_convergence} presents the convergence of the NRMSE as a function of the number of iterations for ePIE, DM, SHARP, and BM-PMACE.
BM-PMACE demonstrates a significantly faster convergence rate and reaches stable solutions within about 50 iterations. 
This indicates the efficiency of BM-PMACE and also its robustness to the presence of noise. 

\begin{figure}[!h]
    \centering
    \includegraphics[width=0.4\textwidth]{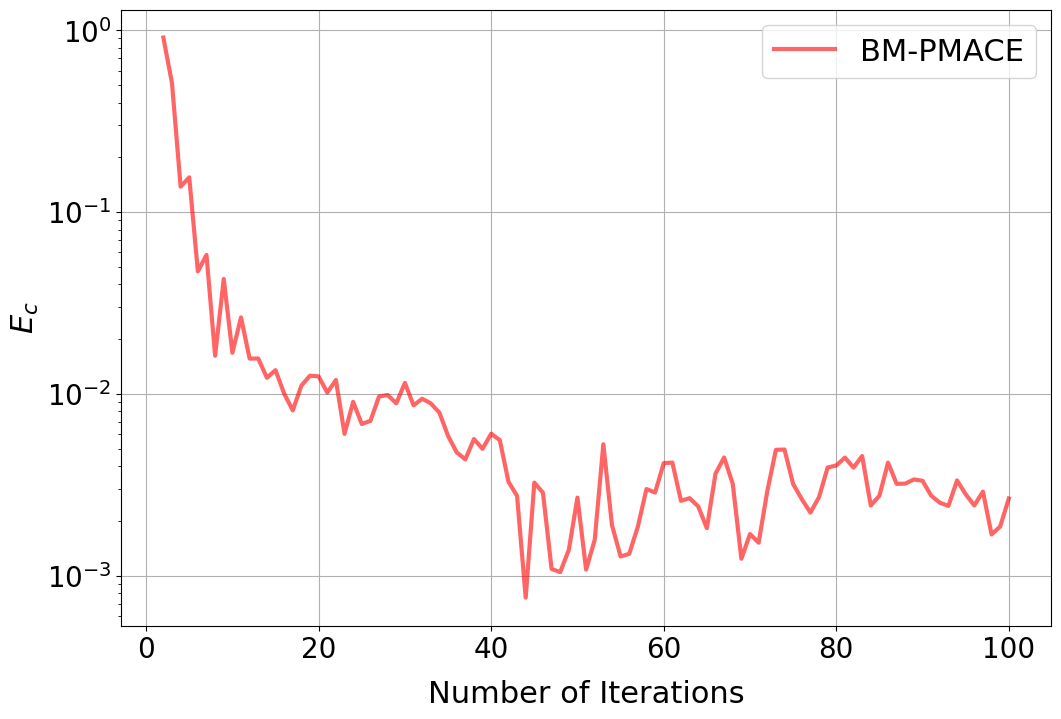}
    \caption{Convergence of patch updates in BM-PMACE as a function of iterations.}
    \label{fig:patch_convergence}
\end{figure}

Figure~\ref{fig:patch_convergence} illustrates the convergence of BM-PMACE as a function of the number of iterations. 
Here, convergence is measured by the quantity $E_c=\frac{1}{J} \left\|  \mathbf{z} - \mathbf{w} \right\|_2$, where $\mathbf{w}$ and $\mathbf{z}$ are variables introduced in Algorithm~\ref{alg: PMACE} and $J$ is the number of patches.
Notice that as $E_c$ approaches zero, the output of $\bF^{I} (\cdot)$ and $\bG^{I}(\cdot)$ become equal, and the desired equilibrium condition of~\eqref{eq:pmace-state-eqns} holds.
Consequently, the plot of Figure~\ref{fig:patch_convergence} indicates that the BM-PMACE algorithm converges to an equilibrium, up to some inherent uncertainty.

\subsection{Multi-Mode Blind Reconstruction: Synthetic Data}

\subsubsection{Multi-Mode Data Simulation}

\begin{figure}[h!]
    \centering
    \includegraphics[width=0.4\textwidth]{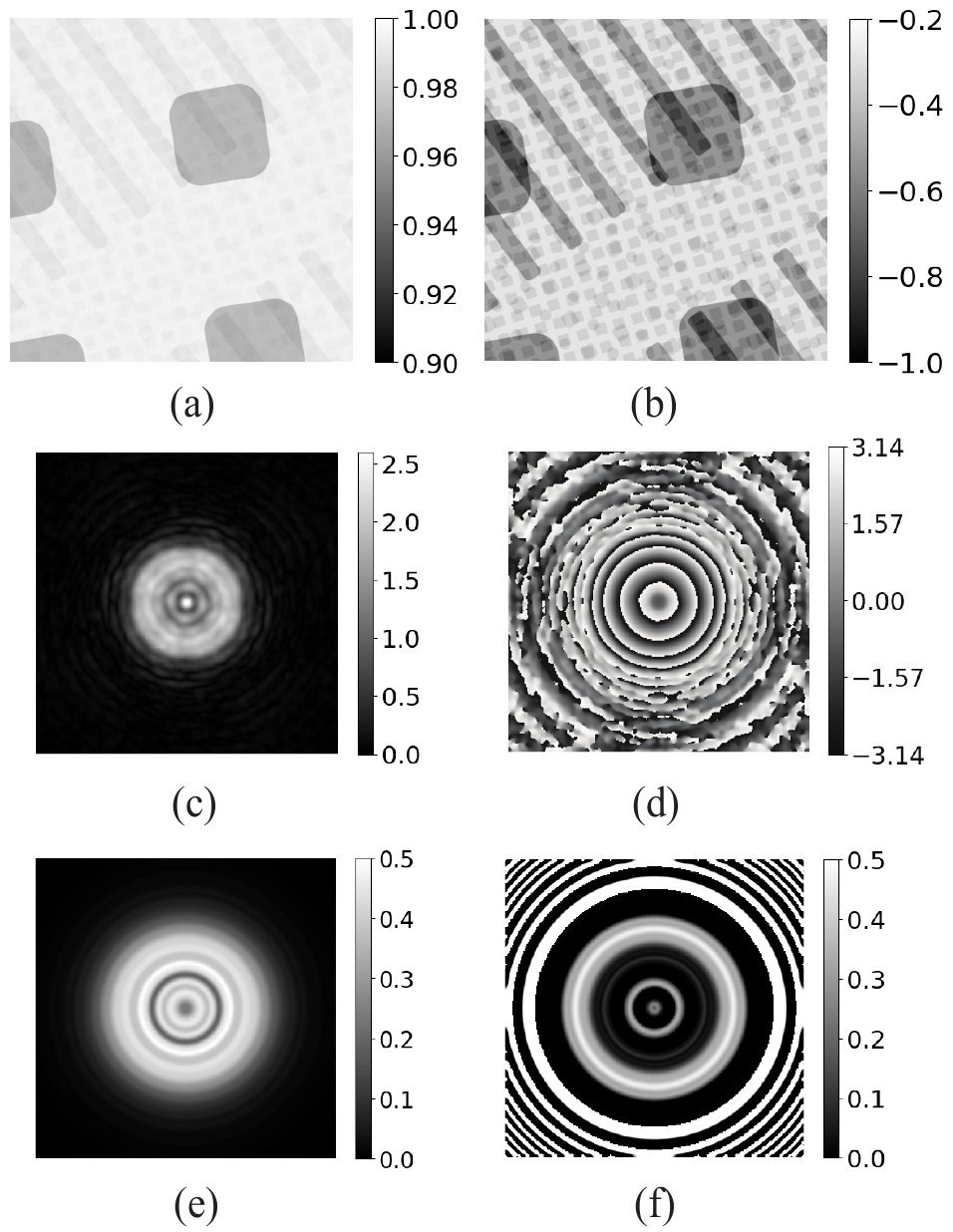}
    \caption{Ground truth image and mutually incoherent probe modes. The complex ground truth object's (a) magnitude and (b) phase; The main probe mode function's (c) magnitude and (d) phase; The secondary probe mode function's (e) magnitude and (f) phase.}
    \label{fig:ground_truth_multi_mode}
\end{figure}

Figure~\ref{fig:ground_truth_multi_mode} shows the $K=2$ probe modes used for the simulation along with the $256 \times 256$ pixel ground truth image.
As with the single-mode experiment, this ground truth image is based on a simulated transmission measurement through a composite material.
The probe modes are $256 \times 256$ pixels and were used in \eqref{eq: synthetic_data_sim} with $K=2$ to generate the simulated data.
As in the single-mode experiment, probe locations were generated on a rectangular grid with randomized offset and an average probe spacing of 36 pixels.
The probe modes were constructed so that the main probe mode occupied 90\% of the total energy, resulting in an overlap ratio of the main probe mode of $r_{ovlp} \approx 65\%$.

\subsubsection{Multi-Mode Reconstruction Results}

We performed reconstruction using ePIE, DM, and BM-PMACE, each for 200 iterations. 
For all methods, the initializations followed our approach, and the secondary probe mode was initialized and introduced to the reconstruction process after 20 iterations.  

\begin{figure}[h!]
    \centering
    \includegraphics[width=0.45\textwidth]{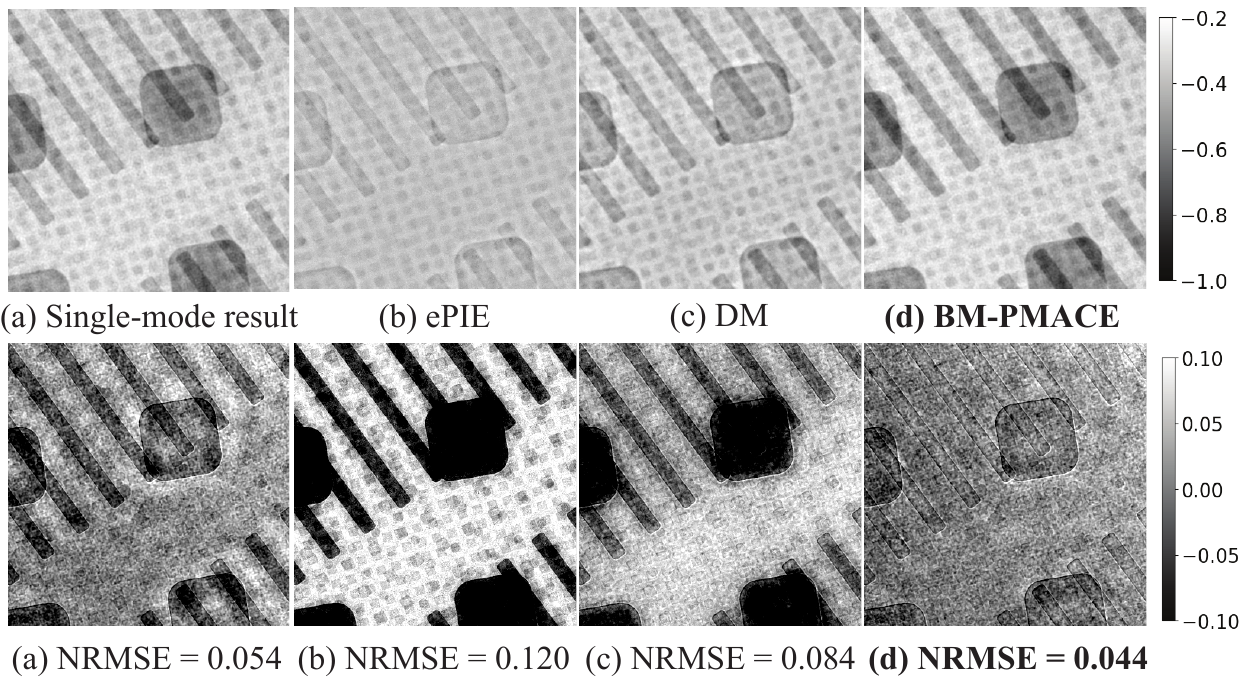}
    \caption{Top:  {\bf Phase} (in radians) of the reconstructed complex transmittance images in Figure~\ref{fig:ground_truth_multi_mode} from synthetic multi-mode data.  Bottom: Difference between the reconstructed and ground truth phase, with NRMSE values indicated in the subcaptions. }
    \label{fig:two_mode_recon_object_phase}
\end{figure}

\begin{figure}[h!]
    \centering
    \includegraphics[width=0.45\textwidth]{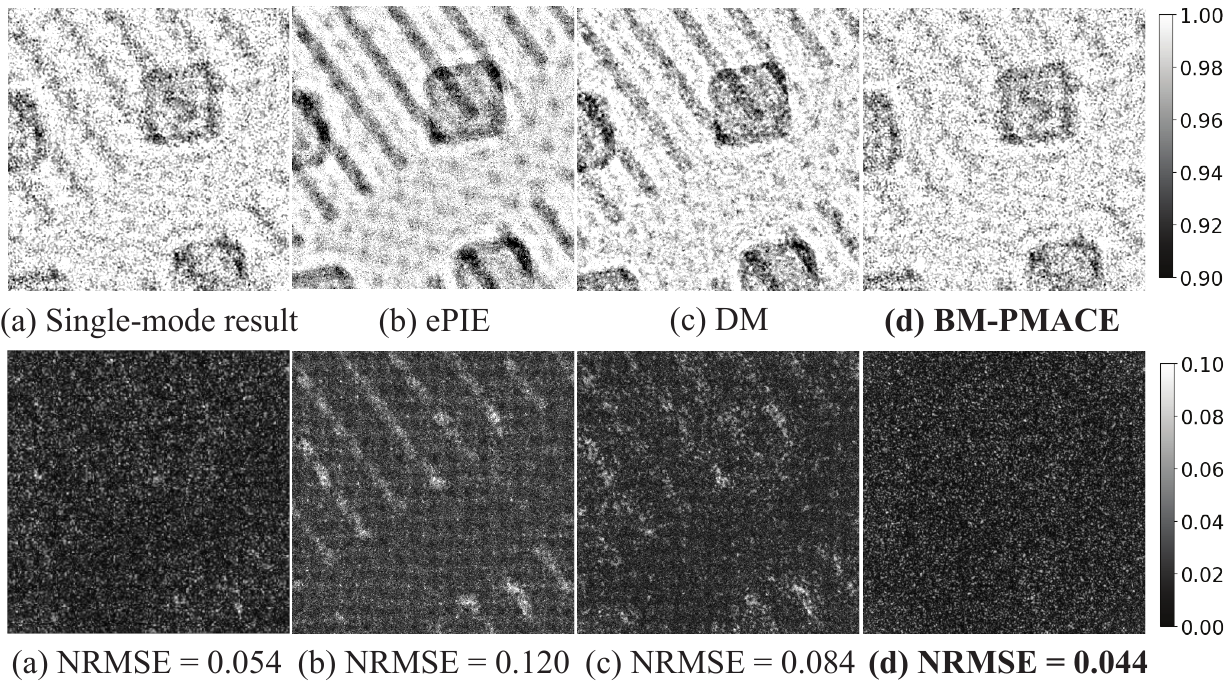}
    \caption{Top:  {\bf Magnitude} of the reconstructed complex transmittance images in Figure~\ref{fig:ground_truth_multi_mode} from synthetic multi-mode data. Bottom: Amplitdues of error between the complex reconstructions and ground truth. }
    \label{fig:two_mode_recon_object_mag}
\end{figure}

\begin{figure}[h!]
    \centering
    \includegraphics[width=0.45\textwidth]{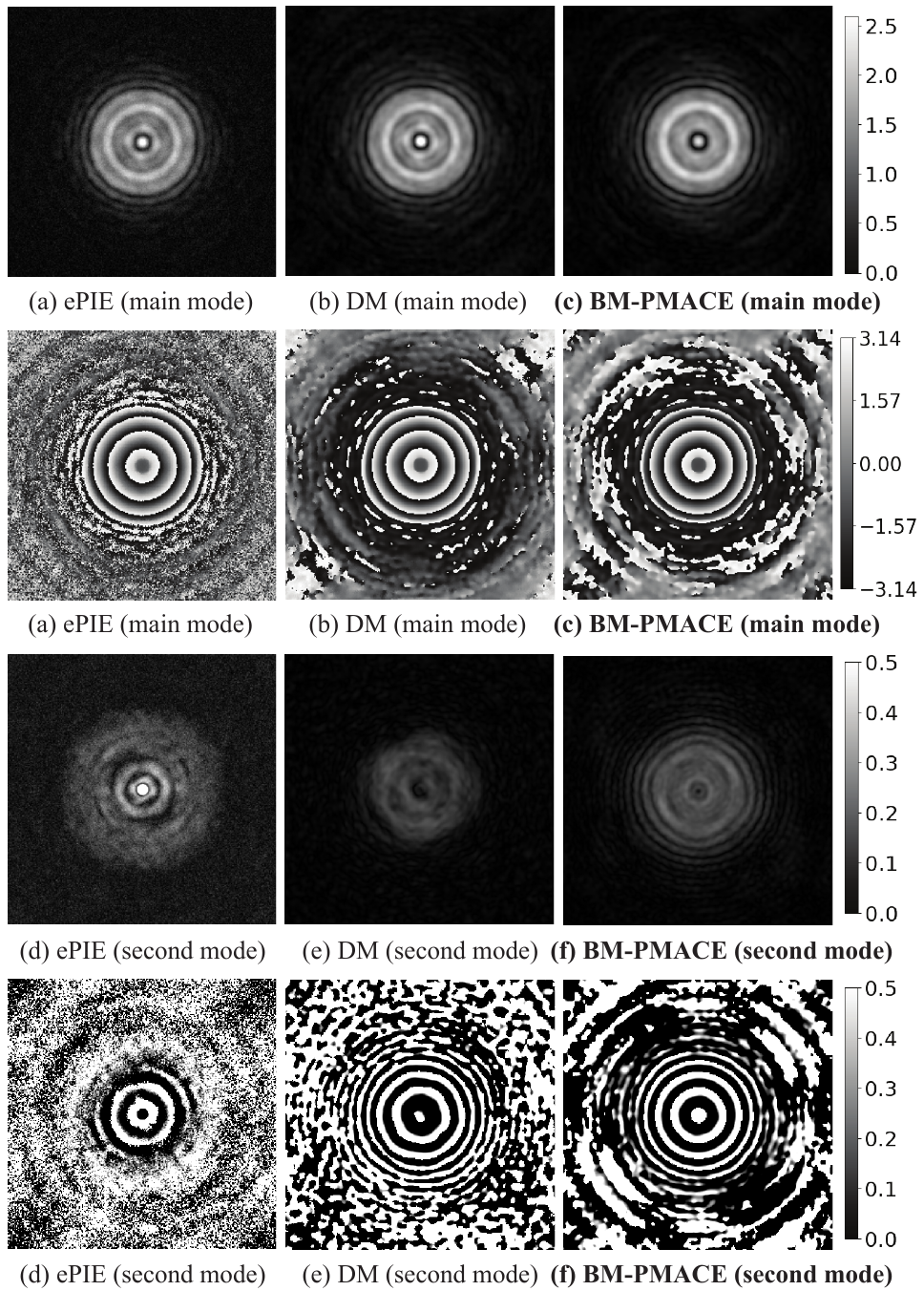}
    \caption{Reconstructed complex probe functions from the multi-mode data. Top row: Magnitudes of the reconstructed complex probe functions for the main mode, using ePIE, DM, and BM-PMACE. Second row: Phase (in radians) of the reconstructed complex probe function for the main mode. Third and fourth rows: Magnitudes and phases (in radians) of the reconstructed complex probe function for the secondary mode from multi-mode data.}
    \label{fig:incoherent_recon_probe}
\end{figure}

Figures~\ref{fig:two_mode_recon_object_phase}
and \ref{fig:two_mode_recon_object_mag} show the reconstructed phases and magnitudes obtained using the ePIE, DM, and BM-PMACE methods. 
Also, the left column in both figures shows the reconstructed images using BM-PMACE with single-mode, which does not fully capture the details in ground truth image. 
Note that the transmittance phase is typically more important than magnitude in many applications. 
This is because phase information often provides much higher contrast than magnitude information. 
In addition, the phase shift is directly related to the optical path length through the object, which provides quantitative information about the sample's thickness and refractive index. 
Our results indicate that BM-PMACE method produces images with significantly fewer artifacts compared to the other approaches and substantially lower NRMSE. 

Figure~\ref{fig:incoherent_recon_probe} shows the reconstructed probe images, including the reconstructed magnitudes and phases of both the main probe mode and the secondary probe mode. 
These figures illustrate the detailed structure of the reconstructed probe modes for each method. 
BM-PMACE consistently captures the complex features of the illumination with greater precision, while ePIE and DM exhibit more distortions and less accurate probe mode representations.

\begin{figure}[!h]
    \centering
    \includegraphics[width=0.45\textwidth]{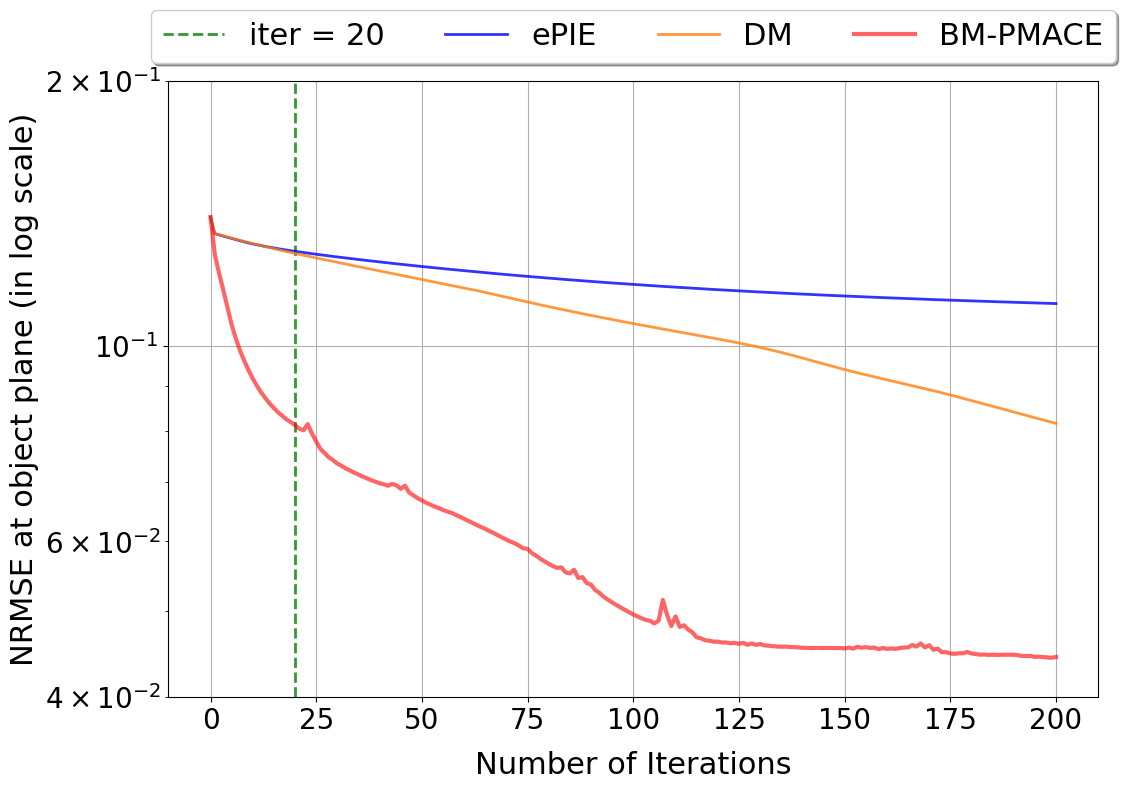}
    \caption{NRMSE between reconstructed object images and ground truth as a function of number of iterations on synthetic multi-mode data.}
    \label{fig:incoherent_convergence}
\end{figure}

Figure~\ref{fig:incoherent_convergence} shows convergence plots for synthetic multi-mode data.
ePIE exhibits fast initial convergence but fails to reach an optimal solution. 
DM converges to a better solution than ePIE but requires significantly more iterations. 
DM continues to improve after 200 iterations, which can result in excessively long computational time. 
BM-PMACE has substantially faster convergence as compared to both ePIE and DM and the solution it reaches has much lower NRMSE than ePIE and DM.

\vspace{-0.2cm}

\begin{figure*}[!htb]
\includegraphics[width=1\textwidth]{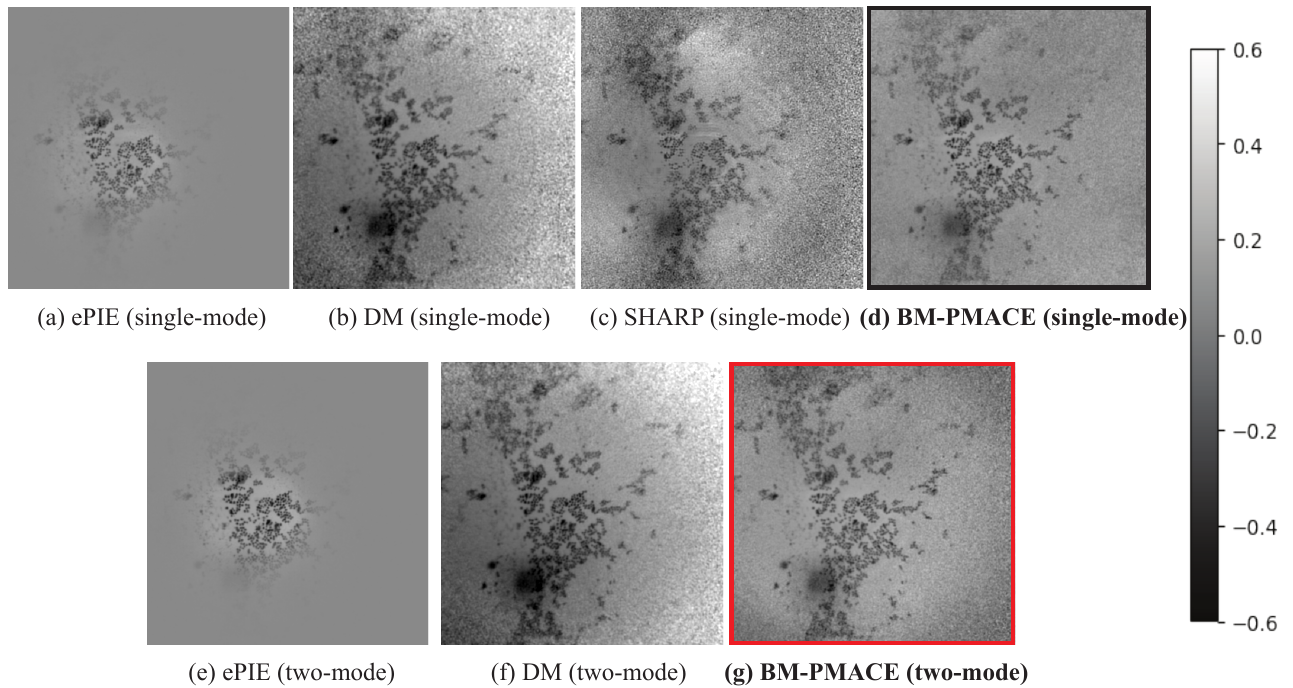}
\caption{{\bf Phase} (in radians) of the reconstructed complex transmittance images from the measured data. The top row features the reconstructed images using ePIE, DM, SHARP, and BM-PMACE, using a single probe mode. The Bottom row shows the reconstructed images using ePIE, DM, and BM-PMACE, using two probe modes.  
}
\vspace{0.2cm}
\label{fig:Auballs_object_phase}
\end{figure*}

\begin{figure*}[!htb]
\includegraphics[width=1\textwidth]{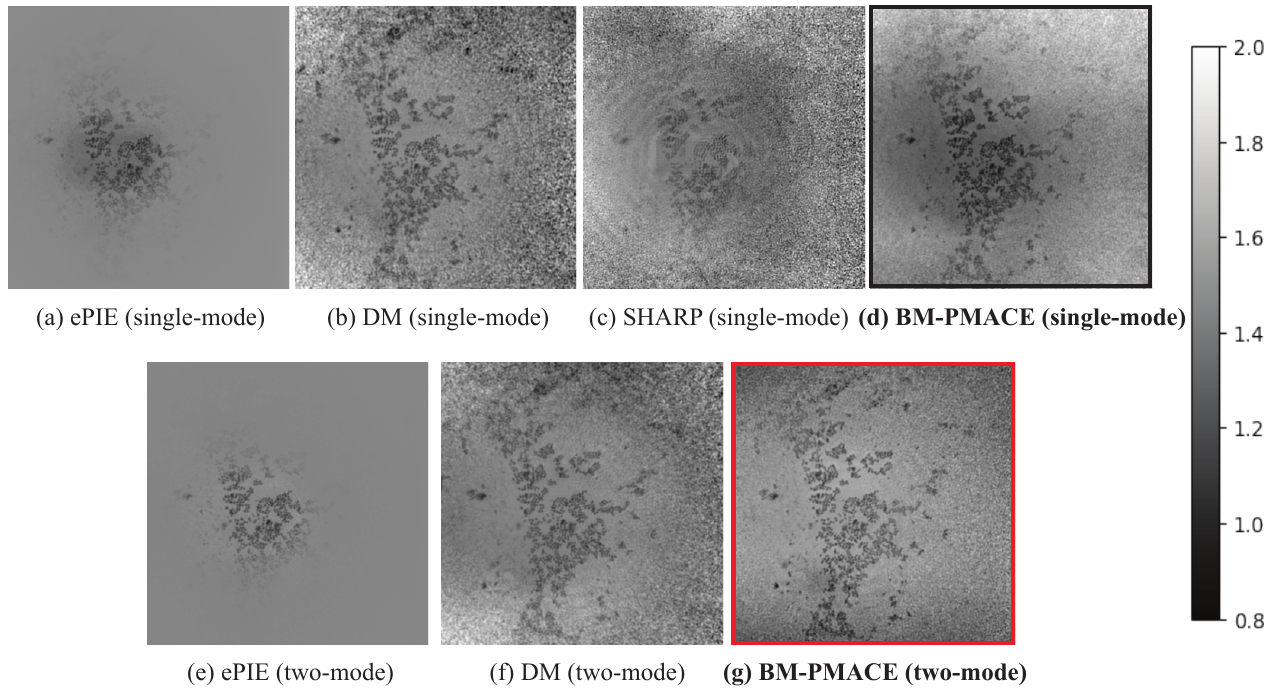}
\caption{{\bf Magnitudes} of the reconstructed complex transmittance images from the measured data. The top row features the reconstructed images using ePIE, DM, SHARP, and BM-PMACE, using a single probe mode. The Bottom row shows the reconstructed images using ePIE, DM, and BM-PMACE, using two probe modes. 
}
\vspace{0.1cm}
\label{fig:Auballs_object_mag}
\end{figure*}

\begin{figure*}[!htb]
\includegraphics[width=1\textwidth]{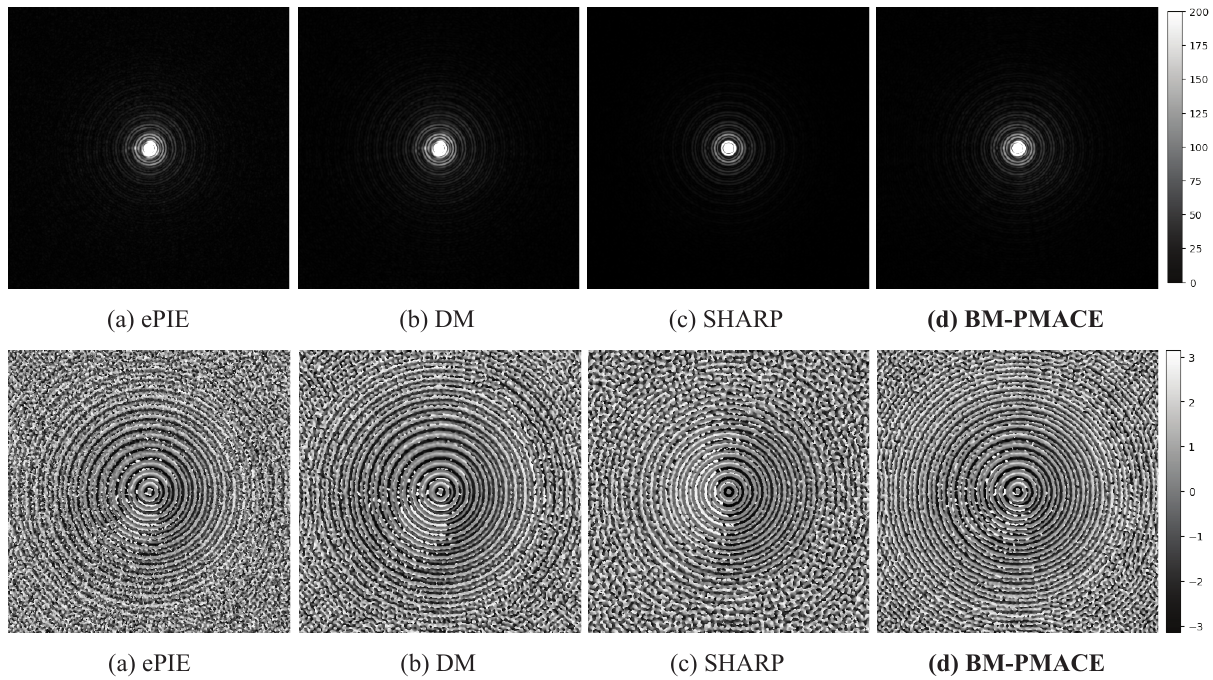}
\caption{Reconstructed complex probe functions from the {\bf single-mode} reconstruction experiment. Top: Magnitudes of reconstructed complex probe functions in single-mode reconstruction using ePIE, DM, SHARP, and BM-PMACE. Bottom: Phase (in radians) of the reconstructed complex probe functions. 
}
\vspace{-0.2cm}
\label{fig:Auballs_probe_single_mode}
\end{figure*}

\begin{figure*}[!htb]
\centering
\includegraphics[width=0.82\textwidth]{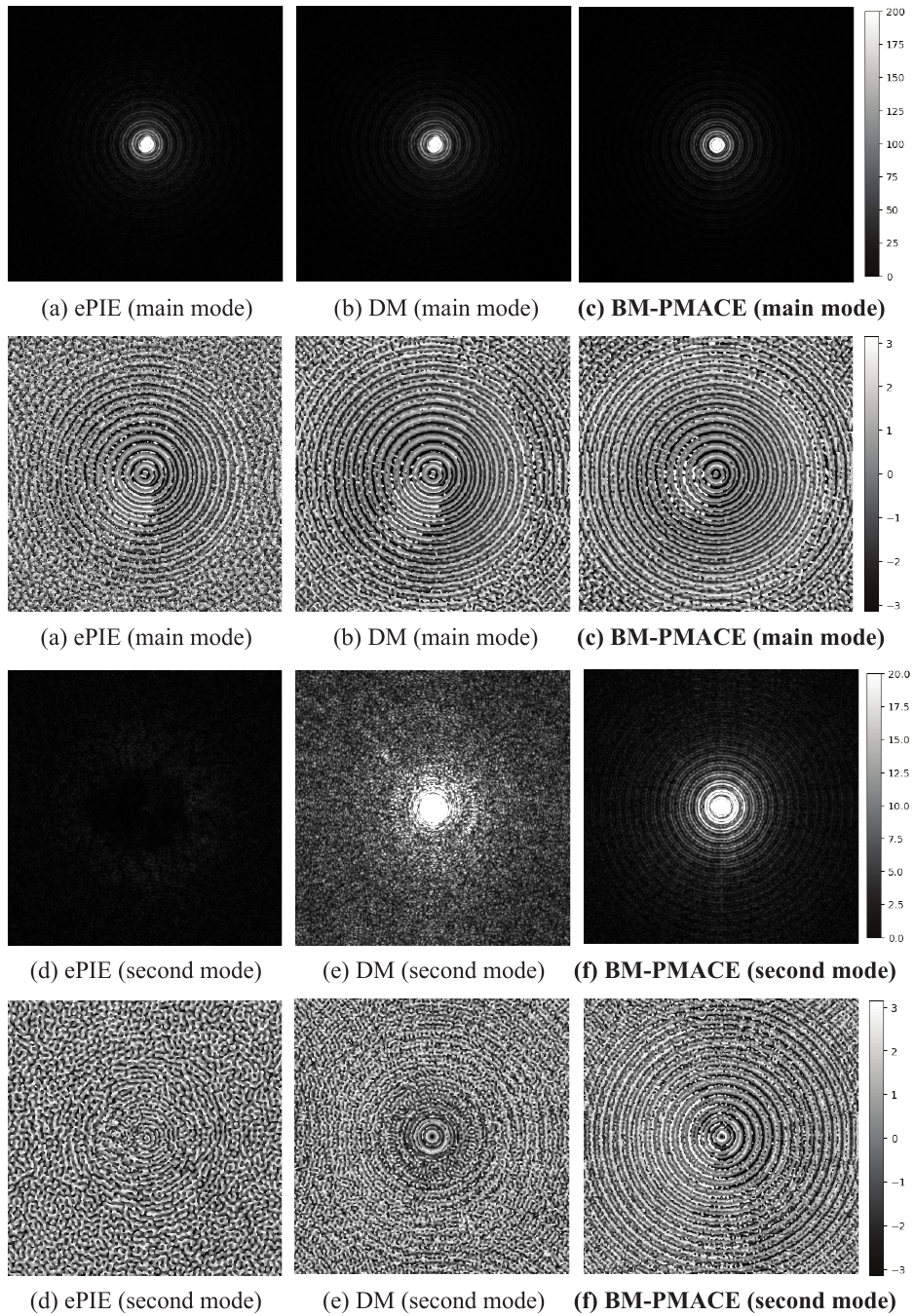}
\vspace{0.2cm}
\caption{Reconstructed complex probe functions from the {\bf two-mode}  reconstruction experiment. Top row: Magnitudes of the reconstructed complex probe functions for the main mode, using ePIE, DM, SHARP, and BM-PMACE. Second row: Phase (in radians) of the reconstructed complex probe function for the main mode. Third and fourth rows: Magnitudes and phases (in radians) of the reconstructed complex probe function for the secondary mode from measured data.
}
\label{fig:Auballs_probe_two_mode}
\end{figure*}

\subsection{Experimental Methods on Measured Data}

Our measured data experiment uses the Ptychography Gold Ball Example Dataset~\cite{osti_1454414} that was collected using the X-ray free-electron lasers with a photon energy level of 1 keV. 
This experiment was conducted at beamline 5.3.2 of the Advanced Light Source (ALS). 
The scans were performed across a $ 20 \times 40$ grid, with each scan being spaced 30 nm apart. 
A total of 800 ptychographic measurements were captured by the detector positioned 0.112 m downstream from the sample of nanometer-sized gold balls. 
Based on prior investigation on this dataset, the illumination overlap ratio is approximately 84\%, which is sufficient for achieving high-quality reconstructions in ptychography \cite{zhai2023projected}.
 
In the preprocessing of the raw data, we subtracted the average of 20 dark scans from each intensity measurement. 
This step is critical for mitigating the influence of noise. Subsequently, a subset of 6 measurements exhibiting significant deviations are identified as outliers and excluded from the dataset. 
This step is crucial for mitigating the impact of inconsistencies from the data and enhancing the fidelity of the reconstruction. 
For fast calculation of Fourier Transform, we cropped each measurement and reduced the dimensions from $621 \times 621$ pixels to $512 \times 512$ pixels. 
While cropping diffraction pattern reduces the maximum scattering angle captured, this is an acceptable trade-off as the cropped data still contains sufficient high-frequency component for the reconstruction.
To further enhance the data quality, we suppress noise by multiplying a 2D Tukey window to each of these diffraction measurements. 
The 2D Tukey window was generated by rotating a 1D Tukey window with shape parameter of 0.5. The resulting data set contains 794 preprocessed and high-quality diffraction measurements. 

We performed reconstructions using single-mode, $K=1$, and two-mode, $K=2$, approaches.
For the single-mode reconstruction, we compared ePIE, DM, SHARP, and BM-PMACE; but for the two-mode reconstruction, we only compared ePIE, DM, and BM-PMACE, since SHARP does not support multi-mode reconstruction.
To quantitatively compare the different methods, we calculated the forward-propagated NRMSE metric in which we evaluated the NRMSE between the measured data and the estimated measurements obtained by passing the reconstruction and estimated probes through the forward model.
Each method was run for a total of 100 iterations and parameters were selected to minimize the forward-propagated NRMSE. 
Initialization was performed as described in Section~\ref{sec:initialization}.

\subsection{Single-Mode and Multi-Mode Results: Measured Data}

Figures~\ref{fig:Auballs_object_phase} and \ref{fig:Auballs_object_mag} show the reconstructed phase and magnitude images for the Gold Ball dataset using single-mode and two-mode approaches.
ePIE achieves good image quality in the central region, but has poor quality across the larger field-of-view. 
SHARP and DM are able to reconstruct features in a relatively larger field of view. 
However, they tend to reconstruct images with blurry details, especially near the edge of reconstructed images. 
BM-PMACE reconstructs the features across the largest field-of-view and exhibits significantly less noise compared to the other methods. 
In general, the two-mode reconstructions are superior to the one-mode reconstructions.

Figure~\ref{fig:Auballs_probe_single_mode} presents the probe estimates using ePIE, DM, SHARP, and BM-PMACE under the coherent probe assumption, where a single probe was reconstructed along with the object transmittance image. 
As the illumination overlap ratio is high and the data is sufficient for reconstruction, all methods produce high-quality probes with reliable and similar density distributions and wavefront shapes. 

Figure~\ref{fig:Auballs_probe_two_mode} shows the reconstructed main mode and secondary mode using ePIE, DM, and BM-PMACE under the multiple probe modes condition, with the secondary probe introduced after 10 iterations. 
All reconstructed main probes capture more than 95\% of the total energy. However, the secondary mode exhibited notable differences. The secondary mode reconstructed using ePIE has low density and carries subtle phase information, and thus appears dim in Figure~\ref{fig:Auballs_probe_two_mode}. 
The secondary mode reconstructed using DM provides more information but still lacks detail. 
BM-PMACE produces a secondary mode with more detailed density and phase information, demonstrating its ability to capture complex features in multi-mode reconstructions.
We limit our experiment to two probe modes, as adding more modes did not further improve reconstruction quality.

\begin{table}[!h]
  \centering
  \begin{tabular}{c c c}
    \toprule
    \multirow{2}{3em}{Method} & \multicolumn{2}{c}{Reconstruction} \\
    & Single-mode & Two-mode \\
    \midrule
    ePIE & 0.090 & 0.070 \\
    DM & 0.082 & 0.069 \\
    SHARP & 0.083 & - \\
    \textbf{BM-PMACE} & \textbf{0.074} & \textbf{0.065} \\
    \bottomrule
  \end{tabular}
  \caption{Comparison of forward-propagated NRMSE for each reconstruction method under single-mode and two-mode conditions. The "-" indicates that the reconstruction condition is not supported by the method.}
  \label{tab:Auballs_detector_NRMSE}
\end{table}

Table~\ref{tab:Auballs_detector_NRMSE} displays the forward-propagated NRMSE values for both single-mode and two-mode cases. 
BM-PMACE achieves the lowest NRMSE of all of the methods in both the single-mode and two-mode cases, indicating that it fits the measurements most effectively. 
Notably, all methods show improvement when an extra mode is incorporated into the reconstruction.

\section{Conclusion}

In this paper we presented BM-PMACE for blind ptychographic reconstruction. Our method offers a robust, efficient, and flexible framework for accurately estimating complex transmittance images and probe functions, leveraging the benefit of local refinement of probe estimate while maintaining the parallel structure among the agents. Additionally, it accommodates multiple probe modes, which enables effective reconstruction in scenarios with partially coherent illumination.

We applied BM-PMACE to blind ptychography on both synthetic noisy data and measured data. The results indicate that BM-PMACE achieves high-quality reconstruction with fewer iterations compared to existing methods, potentially reducing computational time and resources. Additionally, BM-PMACE exhibited rapid convergence to a stable state, even in the presence of noise in data. 
Overall, BM-PMACE provides a reliable framework for accurate and efficient ptychographic reconstruction.

\appendices

\ifCLASSOPTIONcaptionsoff
  \newpage
\fi

\bibliographystyle{IEEEtran}
\bibliography{references}

\begin{thebibliography}{10}
\providecommand{\url}[1]{#1}
\csname url@samestyle\endcsname
\providecommand{\newblock}{\relax}
\providecommand{\bibinfo}[2]{#2}
\providecommand{\BIBentrySTDinterwordspacing}{\spaceskip=0pt\relax}
\providecommand{\BIBentryALTinterwordstretchfactor}{4}
\providecommand{\BIBentryALTinterwordspacing}{\spaceskip=\fontdimen2\font plus
\BIBentryALTinterwordstretchfactor\fontdimen3\font minus \fontdimen4\font\relax}
\providecommand{\BIBforeignlanguage}[2]{{%
\expandafter\ifx\csname l@#1\endcsname\relax
\typeout{** WARNING: IEEEtran.bst: No hyphenation pattern has been}%
\typeout{** loaded for the language `#1'. Using the pattern for}%
\typeout{** the default language instead.}%
\else
\language=\csname l@#1\endcsname
\fi
#2}}
\providecommand{\BIBdecl}{\relax}
\BIBdecl

\bibitem{nellist1995resolution}
P.~Nellist, B.~McCallum, and J.~M. Rodenburg, ``Resolution beyond the'information limit'in transmission electron microscopy,'' \emph{Nature}, vol. 374, no. 6523, pp. 630--632, 1995.

\bibitem{rodenburg2004phase}
J.~M. Rodenburg and H.~M. Faulkner, ``A phase retrieval algorithm for shifting illumination,'' \emph{Applied physics letters}, vol.~85, no.~20, pp. 4795--4797, 2004.

\bibitem{rodenburg2007hard}
J.~M. Rodenburg, A.~Hurst, A.~G. Cullis, B.~R. Dobson, F.~Pfeiffer, O.~Bunk, C.~David, .~f.~K. Jefimovs, and I.~Johnson, ``Hard-x-ray lensless imaging of extended objects,'' \emph{Physical review letters}, vol.~98, no.~3, p. 034801, 2007.

\bibitem{pfeiffer2018x}
F.~Pfeiffer, ``X-ray ptychography,'' \emph{Nature Photonics}, vol.~12, no.~1, pp. 9--17, 2018.

\bibitem{rodenburg2019ptychography}
J.~Rodenburg and A.~Maiden, ``Ptychography,'' \emph{Springer Handbook of Microscopy}, pp. 819--904, 2019.

\bibitem{li2018multi}
P.~Li and A.~Maiden, ``Multi-slice ptychographic tomography,'' \emph{Scientific reports}, vol.~8, no.~1, p. 2049, 2018.

\bibitem{chang2020ptychographic}
D.~J. Chang, D.~S. Kim, A.~Rana, X.~Tian, J.~Zhou, P.~Ercius, and J.~Miao, ``Ptychographic atomic electron tomography: Towards three-dimensional imaging of individual light atoms in materials,'' \emph{Physical Review B}, vol. 102, no.~17, p. 174101, 2020.

\bibitem{batey2022high}
D.~Batey, C.~Rau, and S.~Cipiccia, ``High-speed x-ray ptychographic tomography,'' \emph{Scientific Reports}, vol.~12, no.~1, p. 7846, 2022.

\bibitem{gorecki2023ptychographic}
R.~G{\'o}recki, C.~C. Polo, T.~A. Kalile, E.~X. Miqueles, Y.~R. Tonin, L.~Upadhyaya, F.~Meneau, and S.~P. Nunes, ``Ptychographic x-ray computed tomography of porous membranes with nanoscale resolution,'' \emph{Communications Materials}, vol.~4, no.~1, p.~68, 2023.

\bibitem{pelz2023solving}
P.~M. Pelz, S.~M. Griffin, S.~Stonemeyer, D.~Popple, H.~DeVyldere, P.~Ercius, A.~Zettl, M.~C. Scott, and C.~Ophus, ``Solving complex nanostructures with ptychographic atomic electron tomography,'' \emph{Nature Communications}, vol.~14, no.~1, p. 7906, 2023.

\bibitem{wang2023optical}
T.~Wang, S.~Jiang, P.~Song, R.~Wang, L.~Yang, T.~Zhang, and G.~Zheng, ``Optical ptychography for biomedical imaging: recent progress and future directions,'' \emph{Biomedical Optics Express}, vol.~14, no.~2, pp. 489--532, 2023.

\bibitem{guo2023depth}
C.~Guo, S.~Jiang, L.~Yang, P.~Song, A.~Pirhanov, R.~Wang, T.~Wang, X.~Shao, Q.~Wu, Y.~K. Cho \emph{et~al.}, ``Depth-multiplexed ptychographic microscopy for high-throughput imaging of stacked bio-specimens on a chip,'' \emph{Biosensors and Bioelectronics}, vol. 224, p. 115049, 2023.

\bibitem{wang2017electron}
P.~Wang, F.~Zhang, S.~Gao, M.~Zhang, and A.~I. Kirkland, ``Electron ptychographic diffractive imaging of boron atoms in lab6 crystals,'' \emph{Scientific Reports}, vol.~7, no.~1, p. 2857, 2017.

\bibitem{jiang2022ptychographic}
S.~Jiang, C.~Guo, Z.~Bian, R.~Wang, J.~Zhu, P.~Song, P.~Hu, D.~Hu, Z.~Zhang, K.~Hoshino \emph{et~al.}, ``Ptychographic sensor for large-scale lensless microbial monitoring with high spatiotemporal resolution,'' \emph{Biosensors and Bioelectronics}, vol. 196, p. 113699, 2022.

\bibitem{li20224d}
G.~Li, H.~Zhang, and Y.~Han, ``4d-stem ptychography for electron-beam-sensitive materials,'' \emph{ACS Central Science}, vol.~8, no.~12, pp. 1579--1588, 2022.

\bibitem{chang2018partially}
H.~Chang, P.~Enfedaque, Y.~Lou, and S.~Marchesini, ``Partially coherent ptychography by gradient decomposition of the probe,'' \emph{Acta Crystallographica Section A: Foundations and Advances}, vol.~74, no.~3, pp. 157--169, 2018.

\bibitem{chang2023fast}
H.~Chang, L.~Yang, and S.~Marchesini, ``Fast iterative algorithms for blind phase retrieval: A survey,'' in \emph{Handbook of Mathematical Models and Algorithms in Computer Vision and Imaging: Mathematical Imaging and Vision}.\hskip 1em plus 0.5em minus 0.4em\relax Springer, 2023, pp. 139--174.

\bibitem{deng2015continuous}
J.~Deng, Y.~S. Nashed, S.~Chen, N.~W. Phillips, T.~Peterka, R.~Ross, S.~Vogt, C.~Jacobsen, and D.~J. Vine, ``Continuous motion scan ptychography: characterization for increased speed in coherent x-ray imaging,'' \emph{Optics express}, vol.~23, no.~5, pp. 5438--5451, 2015.

\bibitem{chang2019blind}
H.~Chang, P.~Enfedaque, and S.~Marchesini, ``Blind ptychographic phase retrieval via convergent alternating direction method of multipliers,'' \emph{SIAM Journal on Imaging Sciences}, vol.~12, no.~1, pp. 153--185, 2019.

\bibitem{tamaki2024near}
H.~Tamaki and K.~Saitoh, ``Near-field electron ptychography using full-field structured illumination,'' \emph{Microscopy}, p. dfae035, 2024.

\bibitem{rodenburg2008ptychography}
J.~M. Rodenburg, ``Ptychography and related diffractive imaging methods,'' \emph{Advances in imaging and electron physics}, vol. 150, pp. 87--184, 2008.

\bibitem{moxham2020hard}
T.~E. Moxham, A.~Parsons, T.~Zhou, L.~Alianelli, H.~Wang, D.~Laundy, V.~Dhamgaye, O.~J. Fox, K.~Sawhney, and A.~M. Korsunsky, ``Hard x-ray ptychography for optics characterization using a partially coherent synchrotron source,'' \emph{Journal of Synchrotron Radiation}, vol.~27, no.~6, pp. 1688--1695, 2020.

\bibitem{yao2020multi}
Y.~Yao, Y.~Jiang, J.~A. Klug, M.~Wojcik, E.~R. Maxey, N.~S. Sirica, C.~Roehrig, Z.~Cai, S.~Vogt, B.~Lai \emph{et~al.}, ``Multi-beam x-ray ptychography for high-throughput coherent diffraction imaging,'' \emph{Scientific reports}, vol.~10, no.~1, p. 19550, 2020.

\bibitem{deng2015opportunities}
J.~Deng, D.~J. Vine, S.~Chen, Y.~S. Nashed, T.~Peterka, R.~Ross, S.~Vogt, and C.~J. Jacobsen, ``Opportunities and limitations for combined fly-scan ptychography and fluorescence microscopy,'' in \emph{X-Ray Nanoimaging: Instruments and Methods II}, vol. 9592.\hskip 1em plus 0.5em minus 0.4em\relax SPIE, 2015, pp. 111--119.

\bibitem{huang2015fly}
X.~Huang, K.~Lauer, J.~N. Clark, W.~Xu, E.~Nazaretski, R.~Harder, I.~K. Robinson, and Y.~S. Chu, ``Fly-scan ptychography,'' \emph{Scientific reports}, vol.~5, no.~1, p. 9074, 2015.

\bibitem{pelz2014fly}
P.~M. Pelz, M.~Guizar-Sicairos, P.~Thibault, I.~Johnson, M.~Holler, and A.~Menzel, ``On-the-fly scans for x-ray ptychography,'' \emph{Applied Physics Letters}, vol. 105, no.~25, 2014.

\bibitem{thibault2013reconstructing}
P.~Thibault and A.~Menzel, ``Reconstructing state mixtures from diffraction measurements,'' \emph{Nature}, vol. 494, no. 7435, pp. 68--71, 2013.

\bibitem{batey2014information}
D.~J. Batey, D.~Claus, and J.~M. Rodenburg, ``Information multiplexing in ptychography,'' \emph{Ultramicroscopy}, vol. 138, pp. 13--21, 2014.

\bibitem{fannjiang2020blind}
A.~Fannjiang and P.~Chen, ``Blind ptychography: uniqueness and ambiguities,'' \emph{Inverse Problems}, vol.~36, no.~4, p. 045005, 2020.

\bibitem{hesse2015proximal}
R.~Hesse, D.~R. Luke, S.~Sabach, and M.~K. Tam, ``Proximal heterogeneous block implicit-explicit method and application to blind ptychographic diffraction imaging,'' \emph{SIAM Journal on Imaging Sciences}, vol.~8, no.~1, pp. 426--457, 2015.

\bibitem{faulkner2004movable}
H.~M.~L. Faulkner and J.~Rodenburg, ``Movable aperture lensless transmission microscopy: a novel phase retrieval algorithm,'' \emph{Physical review letters}, vol.~93, no.~2, p. 023903, 2004.

\bibitem{maiden2009improved}
A.~M. Maiden and J.~M. Rodenburg, ``An improved ptychographical phase retrieval algorithm for diffractive imaging,'' \emph{Ultramicroscopy}, vol. 109, no.~10, pp. 1256--1262, 2009.

\bibitem{maiden2012ptychographic}
A.~M. Maiden, M.~J. Humphry, and J.~M. Rodenburg, ``Ptychographic transmission microscopy in three dimensions using a multi-slice approach,'' \emph{JOSA A}, vol.~29, no.~8, pp. 1606--1614, 2012.

\bibitem{maiden2017further}
A.~Maiden, D.~Johnson, and P.~Li, ``Further improvements to the ptychographical iterative engine,'' \emph{Optica}, vol.~4, no.~7, pp. 736--745, 2017.

\bibitem{thibault2008high}
P.~Thibault, M.~Dierolf, A.~Menzel, O.~Bunk, C.~David, and F.~Pfeiffer, ``High-resolution scanning x-ray diffraction microscopy,'' \emph{Science}, vol. 321, no. 5887, pp. 379--382, 2008.

\bibitem{thibault2009probe}
P.~Thibault, M.~Dierolf, O.~Bunk, A.~Menzel, and F.~Pfeiffer, ``Probe retrieval in ptychographic coherent diffractive imaging,'' \emph{Ultramicroscopy}, vol. 109, no.~4, pp. 338--343, 2009.

\bibitem{marchesini2013augmented}
S.~Marchesini, A.~Schirotzek, C.~Yang, H.-t. Wu, and F.~Maia, ``Augmented projections for ptychographic imaging,'' \emph{Inverse Problems}, vol.~29, no.~11, p. 115009, 2013.

\bibitem{marchesini2016sharp}
S.~Marchesini, H.~Krishnan, B.~J. Daurer, D.~A. Shapiro, T.~Perciano, J.~A. Sethian, and F.~R. Maia, ``{SHARP}: a distributed {GPU}-based ptychographic solver,'' \emph{Journal of Applied Crystallography}, vol.~49, no.~4, pp. 1245--1252, 2016.

\bibitem{yao2021broadband}
Y.~Yao, Y.~Jiang, J.~Klug, Y.~Nashed, C.~Roehrig, C.~Preissner, F.~Marin, M.~Wojcik, O.~Cossairt, Z.~Cai \emph{et~al.}, ``Broadband x-ray ptychography using multi-wavelength algorithm,'' \emph{Journal of Synchrotron Radiation}, vol.~28, no.~1, pp. 309--317, 2021.

\bibitem{long2022single}
X.~Long, L.~Rong, H.~Lin, Y.~Wang, J.~Zhao, S.~Lin, and D.~Wang, ``Single-shot ptychography based on spatial light modulator multi-angle modulation,'' in \emph{Thirteenth International Conference on Information Optics and Photonics (CIOP 2022)}, vol. 12478.\hskip 1em plus 0.5em minus 0.4em\relax SPIE, 2022, pp. 909--918.

\bibitem{dong2018high}
Z.~Dong, Y.-L.~L. Fang, X.~Huang, H.~Yan, S.~Ha, W.~Xu, Y.~S. Chu, S.~I. Campbell, and M.~Lin, ``High-performance multi-mode ptychography reconstruction on distributed gpus,'' in \emph{2018 New York Scientific Data Summit (NYSDS)}.\hskip 1em plus 0.5em minus 0.4em\relax IEEE, 2018, pp. 1--5.

\bibitem{shi2018multi}
X.~Shi, N.~Burdet, D.~Batey, and I.~Robinson, ``Multi-modal ptychography: Recent developments and applications,'' \emph{Applied Sciences}, vol.~8, no.~7, p. 1054, 2018.

\bibitem{fang2020accelerated}
Y.-L.~L. Fang, S.~Ha, X.~Huang, H.~Yan, Z.~Dong, Y.~S. Chu, S.~I. Campbell, W.~Xu, and M.~Lin, ``Accelerated computing for x-ray ptychography at nsls-ii,'' in \emph{Handbook on Big Data and Machine Learning in the Physical Sciences: Volume 2. Advanced Analysis Solutions for Leading Experimental Techniques}.\hskip 1em plus 0.5em minus 0.4em\relax World Scientific, 2020, pp. 141--157.

\bibitem{yu2022scalable}
X.~Yu, V.~Nikitin, D.~J. Ching, S.~Aslan, D.~G{\"u}rsoy, and T.~Bi{\c{c}}er, ``Scalable and accurate multi-{GPU}-based image reconstruction of large-scale ptychography data,'' \emph{Scientific reports}, vol.~12, no.~1, p. 5334, 2022.

\bibitem{bunk2008influence}
O.~Bunk, M.~Dierolf, S.~Kynde, I.~Johnson, O.~Marti, and F.~Pfeiffer, ``Influence of the overlap parameter on the convergence of the ptychographical iterative engine,'' \emph{Ultramicroscopy}, vol. 108, no.~5, pp. 481--487, 2008.

\bibitem{melnyk2023convergence}
O.~Melnyk, ``Convergence properties of gradient methods for blind ptychography,'' \emph{arXiv preprint arXiv:2306.08750}, 2023.

\bibitem{elser2003phase}
V.~Elser, ``Phase retrieval by iterated projections,'' \emph{JOSA A}, vol.~20, no.~1, pp. 40--55, 2003.

\bibitem{xu2018accelerated}
R.~Xu, M.~Soltanolkotabi, J.~P. Haldar, W.~Unglaub, J.~Zusman, A.~F. Levi, and R.~M. Leahy, ``Accelerated {W}irtinger flow: A fast algorithm for ptychography,'' \emph{arXiv preprint arXiv:1806.05546}, 2018.

\bibitem{enfedaque2019high}
P.~Enfedaque, H.~Chang, B.~Enders, D.~Shapiro, and S.~Marchesini, ``High performance partial coherent x-ray ptychography,'' in \emph{Computational Science--ICCS 2019}.\hskip 1em plus 0.5em minus 0.4em\relax Springer, Jun. 2019, pp. 46--59.

\bibitem{zhai2023projected}
Q.~Zhai, G.~T. Buzzard, K.~Mertes, B.~Wohlberg, and C.~A. Bouman, ``Projected multi-agent consensus equilibrium ({PMACE}) with application to ptychography,'' \emph{IEEE Transactions on Computational Imaging}, 2023.

\bibitem{nashed2014parallel}
Y.~S. Nashed, D.~J. Vine, T.~Peterka, J.~Deng, R.~Ross, and C.~Jacobsen, ``Parallel ptychographic reconstruction,'' \emph{Optics express}, vol.~22, no.~26, pp. 32\,082--32\,097, 2014.

\bibitem{zhai2021projected}
Q.~Zhai, B.~Wohlberg, G.~T. Buzzard, and C.~A. Bouman, ``Projected multi-agent consensus equilibrium for ptychographic image reconstruction,'' in \emph{2021 55th Asilomar Conference on Signals, Systems, and Computers}.\hskip 1em plus 0.5em minus 0.4em\relax IEEE, 2021, pp. 1694--1698.

\bibitem{bian2016fourier}
L.~Bian, J.~Suo, J.~Chung, X.~Ou, C.~Yang, F.~Chen, and Q.~Dai, ``Fourier ptychographic reconstruction using poisson maximum likelihood and truncated {W}irtinger gradient,'' \emph{Scientific reports}, vol.~6, no.~1, p. 27384, 2016.

\bibitem{godard2012noise}
P.~Godard, M.~Allain, V.~Chamard, and J.~Rodenburg, ``Noise models for low counting rate coherent diffraction imaging,'' \emph{Optics express}, vol.~20, no.~23, pp. 25\,914--25\,934, 2012.

\bibitem{osti_1454414}
\BIBentryALTinterwordspacing
S.~Marchesini, ``Ptychography gold ball example dataset,'' 7 2017. [Online]. Available: \url{https://www.osti.gov/biblio/1454414}
\BIBentrySTDinterwordspacing

\end{thebibliography}

\end{document}